\documentclass[aps,prd,twocolumn, superscriptaddress, showkeys]{revtex4-1}

\usepackage{multirow}
\usepackage{graphicx}  
\usepackage{dcolumn}   
\usepackage{bm}        
\usepackage{amssymb}   
\usepackage{graphicx, epsfig, subfigure}
\usepackage{hyperref}
\usepackage[mathlines]{lineno}
\usepackage{color}
\usepackage{float}
\usepackage{enumerate}
\usepackage{enumitem}
\usepackage{tabularx}
\usepackage{multirow}
\usepackage{amsmath}
\usepackage{tensor}
\usepackage[export]{adjustbox}
\usepackage{upgreek} 

\usepackage{numcompress}
\newcommand{\invpb}{{pb}$^{-1}$}
\newcommand{\gev}{GeV/$c$}
\newcommand{\gevtwo}{GeV/$c^{2}$}

\newcommand{\jpsi}{\ensuremath{J/\psi}}
\newcommand{\pp}{$p$+$p$}
\newcommand{\sqrts}{\ensuremath{\sqrt{s}}}

\newcommand{\pT}{\ensuremath{p_\mathrm{T}}}
\newcommand{\ltheta}{\ensuremath{\uplambda_{\rm{\uptheta}}}}
\newcommand{\lphi}{\ensuremath{\uplambda_{\rm{\upphi}}}}
\newcommand{\lthetaphi}{\ensuremath{\uplambda_{\rm{\uptheta\upphi}}}}
\newcommand{\linv}{\ensuremath{\uplambda_{\rm{inv}}}}
\newcommand{\nsigmapi}{\ensuremath{\mathrm{n}\sigma_{\pi}}}
\newcommand{\nsigmae}{\ensuremath{\mathrm{n}\sigma_{e}}}
\newcommand{\deltaz}{\ensuremath{\Delta z}}
\newcommand{\deltay}{\ensuremath{\Delta y}}
\newcommand{\timeofflight}{\ensuremath{t_{\rm{tof}}}}	
\newcommand{\deltatof}{\ensuremath{\Delta \timeofflight}}
\newcommand{\DetAccEff}{\ensuremath{A\times \varepsilon}}

\newcommand{\tabincell}[2]{\begin{tabular}{@{}#1@{}}#2\end{tabular}}

\begin{document}

    \title{ Measurement of inclusive \jpsi\ polarization in \pp\ collisions at \sqrts = 200 GeV by the STAR experiment} 

%
%
\affiliation{Abilene Christian University, Abilene, Texas   79699}
\affiliation{AGH University of Science and Technology, FPACS, Cracow 30-059, Poland}
\affiliation{Alikhanov Institute for Theoretical and Experimental Physics NRC "Kurchatov Institute", Moscow 117218, Russia}
\affiliation{Argonne National Laboratory, Argonne, Illinois 60439}
\affiliation{American University of Cairo, New Cairo 11835, New Cairo, Egypt}
\affiliation{Brookhaven National Laboratory, Upton, New York 11973}
\affiliation{University of California, Berkeley, California 94720}
\affiliation{University of California, Davis, California 95616}
\affiliation{University of California, Los Angeles, California 90095}
\affiliation{University of California, Riverside, California 92521}
\affiliation{Central China Normal University, Wuhan, Hubei 430079 }
\affiliation{University of Illinois at Chicago, Chicago, Illinois 60607}
\affiliation{Creighton University, Omaha, Nebraska 68178}
\affiliation{Czech Technical University in Prague, FNSPE, Prague 115 19, Czech Republic}
\affiliation{Technische Universit\"at Darmstadt, Darmstadt 64289, Germany}
\affiliation{ELTE E\"otv\"os Lor\'and University, Budapest, Hungary H-1117}
\affiliation{Frankfurt Institute for Advanced Studies FIAS, Frankfurt 60438, Germany}
\affiliation{Fudan University, Shanghai, 200433 }
\affiliation{University of Heidelberg, Heidelberg 69120, Germany }
\affiliation{University of Houston, Houston, Texas 77204}
\affiliation{Huzhou University, Huzhou, Zhejiang  313000}
\affiliation{Indian Institute of Science Education and Research (IISER), Berhampur 760010 , India}
\affiliation{Indian Institute of Science Education and Research (IISER) Tirupati, Tirupati 517507, India}
\affiliation{Indian Institute Technology, Patna, Bihar 801106, India}
\affiliation{Indiana University, Bloomington, Indiana 47408}
\affiliation{Institute of Modern Physics, Chinese Academy of Sciences, Lanzhou, Gansu 730000 }
\affiliation{University of Jammu, Jammu 180001, India}
\affiliation{Joint Institute for Nuclear Research, Dubna 141 980, Russia}
\affiliation{Kent State University, Kent, Ohio 44242}
\affiliation{University of Kentucky, Lexington, Kentucky 40506-0055}
\affiliation{Lawrence Berkeley National Laboratory, Berkeley, California 94720}
\affiliation{Lehigh University, Bethlehem, Pennsylvania 18015}
\affiliation{Max-Planck-Institut f\"ur Physik, Munich 80805, Germany}
\affiliation{Michigan State University, East Lansing, Michigan 48824}
\affiliation{National Research Nuclear University MEPhI, Moscow 115409, Russia}
\affiliation{National Institute of Science Education and Research, HBNI, Jatni 752050, India}
\affiliation{National Cheng Kung University, Tainan 70101 }
\affiliation{Nuclear Physics Institute of the CAS, Rez 250 68, Czech Republic}
\affiliation{Ohio State University, Columbus, Ohio 43210}
\affiliation{Old Dominion University, Norfolk, VA 23529}
\affiliation{Institute of Nuclear Physics PAN, Cracow 31-342, Poland}
\affiliation{Panjab University, Chandigarh 160014, India}
\affiliation{Pennsylvania State University, University Park, Pennsylvania 16802}
\affiliation{NRC "Kurchatov Institute", Institute of High Energy Physics, Protvino 142281, Russia}
\affiliation{Purdue University, West Lafayette, Indiana 47907}
\affiliation{Rice University, Houston, Texas 77251}
\affiliation{Rutgers University, Piscataway, New Jersey 08854}
\affiliation{Universidade de S\~ao Paulo, S\~ao Paulo, Brazil 05314-970}
\affiliation{University of Science and Technology of China, Hefei, Anhui 230026}
\affiliation{Shandong University, Qingdao, Shandong 266237}
\affiliation{Shanghai Institute of Applied Physics, Chinese Academy of Sciences, Shanghai 201800}
\affiliation{Southern Connecticut State University, New Haven, Connecticut 06515}
\affiliation{State University of New York, Stony Brook, New York 11794}
\affiliation{Temple University, Philadelphia, Pennsylvania 19122}
\affiliation{Texas A\&M University, College Station, Texas 77843}
\affiliation{University of Texas, Austin, Texas 78712}
\affiliation{Tsinghua University, Beijing 100084}
\affiliation{University of Tsukuba, Tsukuba, Ibaraki 305-8571, Japan}
\affiliation{United States Naval Academy, Annapolis, Maryland 21402}
\affiliation{Valparaiso University, Valparaiso, Indiana 46383}
\affiliation{Variable Energy Cyclotron Centre, Kolkata 700064, India}
\affiliation{Warsaw University of Technology, Warsaw 00-661, Poland}
\affiliation{Wayne State University, Detroit, Michigan 48201}
\affiliation{Yale University, New Haven, Connecticut 06520}

\author{J.~Adam}\affiliation{Brookhaven National Laboratory, Upton, New York 11973}
\author{L.~Adamczyk}\affiliation{AGH University of Science and Technology, FPACS, Cracow 30-059, Poland}
\author{J.~R.~Adams}\affiliation{Ohio State University, Columbus, Ohio 43210}
\author{J.~K.~Adkins}\affiliation{University of Kentucky, Lexington, Kentucky 40506-0055}
\author{G.~Agakishiev}\affiliation{Joint Institute for Nuclear Research, Dubna 141 980, Russia}
\author{M.~M.~Aggarwal}\affiliation{Panjab University, Chandigarh 160014, India}
\author{Z.~Ahammed}\affiliation{Variable Energy Cyclotron Centre, Kolkata 700064, India}
\author{I.~Alekseev}\affiliation{Alikhanov Institute for Theoretical and Experimental Physics NRC "Kurchatov Institute", Moscow 117218, Russia}\affiliation{National Research Nuclear University MEPhI, Moscow 115409, Russia}
\author{D.~M.~Anderson}\affiliation{Texas A\&M University, College Station, Texas 77843}
\author{A.~Aparin}\affiliation{Joint Institute for Nuclear Research, Dubna 141 980, Russia}
% ...

\collaboration{STAR Collaboration}\noaffiliation

\date{\today}

\begin{abstract}
We report on new measurements of inclusive \jpsi\ polarization at mid-rapidity in \pp\ collisions at \sqrts\ = 200 GeV by the STAR experiment at RHIC. The polarization parameters, \ltheta, \lphi, and \lthetaphi, are measured as a function of transverse momentum (\pT) in both the Helicity and Collins-Soper (CS) reference frames within $ \pT < 10$ \gev. Except for \ltheta\ in the CS frame at the highest measured \pT, all three polarization parameters are consistent with 0 in both reference frames without any strong \pT\ dependence. Several model calculations are compared with data, and the one using the Color Glass Condensate effective field theory coupled with non-relativistic QCD gives the best overall description of the experimental results, even though other models cannot be ruled out due to experimental uncertainties.
\end{abstract}

\keywords{STAR, \jpsi\ polarization}

\maketitle


\section{Introduction}
The \jpsi\ meson, a bound state of a charm ($c$) and an anti-charm ($\bar{c}$) quark, provides a natural testing ground for studying both the perturbative and non-perturbative aspects of the Quantum Chromodynamics (QCD). Due to their large masses, the production cross section of $c\bar{c}$ pairs can be calculated perturbatively. 
On the other hand, the formation of \jpsi\ mesons from $c\bar{c}$ pairs happens over long distances, and therefore is non-perturbative. The \jpsi\ mesons are also widely used in heavy-ion physics as an internal probe to study the properties of the quark-gluon plasma \cite{Adam:2019rbk}, which requires the measurement of the \jpsi\ production in vacuum as a reference. Despite decades of concentrated experimental and theoretical efforts, a complete picture of the \jpsi\ production mechanism in elementary collisions has yet to emerge. 

Model calculations describing the \jpsi\ production utilize the factorization of the short-distance $c\bar{c}$ production and the long-distance hadronization process \cite{lansberg2019new}. Models differ mainly in the treatment of the non-perturbative formation of \jpsi. One of the early models is the Color Evaporation Model (CEM) \cite{Fritzsch:1977ay, Halzen:1977rs}, which is based on the principle of quark-hadron duality and satisfies all-order factorization. It assumes that every $c\bar{c}$ pair, with an invariant mass below twice the $D$-meson threshold, evolves into a \jpsi\ meson with a fixed probability ($F_{\jpsi}$) by randomly emitting or exchanging soft gluons with other color sources. The non-perturbative \jpsi\ formation is incorporated into the universal probability $F_{\jpsi}$, which is independent of the kinematics and spin of the \jpsi\ meson. An Improved Color Evaporation Model (ICEM) has recently been proposed, in which the lower limit of the $c\bar{c}$ pair invariant mass is increased to be the charmonium mass and the transverse momentum (\pT) of the charmonium state is adjusted based on the ratio of its mass to the $c\bar{c}$ mass \cite{Ma:2016exq}. The ICEM calculation is in general agreement with the inclusive \jpsi\ cross section measured in \pp\ collisions at \sqrts\ = 200 GeV \cite{Ma:2016exq},  in which the discrepancy seen above \jpsi\ \pT\ $\sim$ 4 \gev\ is mainly due to the missing contribution of $b$-hadron decays in ICEM.
By summing with the contribution of \jpsi\ from $b$-hadron decays obtained from the fixed-order plus next-to-leading logarithm (FONLL) calculation \cite{Cacciari:1998it}, the ICEM calculation agrees reasonably well with the inclusive \jpsi\ cross section measured in \pp\ collisions at \sqrts\ = 500 GeV \cite{Adam:2019mrg} up to $\pT^{\jpsi}=20$ \gev. A further extension based on ICEM at leading order (LO) is the calculation of \jpsi\ polarization utilizing the $k_{T}$-factorization approach \cite{Cheung:2018tvq}. Compared to the measured \jpsi\ polarization at forward rapidity in \pp\ collisions at \sqrts\ = 7 TeV \cite{Abelev:2011md, Aaij:2013nlm}, the ICEM calculation shows significant discrepancies at low \pT. 

A more sophisticated way to describe the hadronization of heavy quarkonia is based on the effective quantum field theory of non-relativistic QCD (NRQCD) \cite{Bodwin:1994jh}. In addition to the usual expansion in the strong coupling constant ($\alpha_{s}$), it also introduces an expansion in the relative velocity between the heavy quarks in the pair. Both the color-singlet and color-octet intermediate $c\bar{c}$ pairs are included in the NRQCD. The hadronization process is incorporated through the assumed universal Long Distance Matrix Elements (LDMEs), which weight the relative contributions of each intermediate state and are extracted from fitting experimental data. The NRQCD calculations at next-to-leading order (NLO) in $\alpha_{s}$ have been done by three groups \cite{Butenschoen:2012px,Chao:2012iv,Gong:2012ug}. They obtained very different LDMEs depending on the low-\pT\ cuts imposed on data points used and whether the polarization data are included. None of these calculations can give a simultaneous description of both the charmonium cross section, such as the $\eta_{c}$ yields measured in 7 TeV \pp\ collisions \cite{Aaij:2014bga}, and polarization such as those measured by the CDF Collaboration \cite{Affolder:2000nn, Abulencia:2007us}. To remedy the issue of calculating the $c\bar{c}$ production cross section at low \pT, where the collinear factorization formalism may not be applicable, an effort has been made to use the Color Glass Condensate (CGC) effective field theory \cite{Iancu:2003xm}. Combined with the NRQCD, it describes well the \jpsi\ cross sections measured in \pp\ collisions at both RHIC and the LHC \cite{Ma:2014mri}. The CGC+NRQCD formalism has also been used to calculate the \jpsi\ polarization and the results agree well with the LHC measurements at forward rapidities \cite{Ma:2018qvc}. Continued efforts from both experimental and theoretical sides are still needed to achieve the final goal of a complete understanding of \jpsi\ production. 

While the \jpsi\ production cross section has been measured extensively in \pp\ collisions at \sqrts\ = 200 GeV at RHIC \cite{Adare:2009js,Adamczyk:2012ey,Adam:2018jmp}, its polarization, which is the topic of this paper, is less so \cite{Adare:2009js,Adamczyk:2013vjy}. The \jpsi\ polarization can be measured through the angular distribution of the positively charged daughter lepton \cite{Noman:1978eh}:
\begin{equation}
\begin{split}
W(\cos\uptheta,\upphi) \propto \frac{1}{3+\ltheta} & (1+\ltheta\cos^{2}\uptheta+\lphi\sin^{2}\uptheta \cos2\upphi \\ 
& +\lthetaphi\sin2\uptheta \cos\upphi),
\end{split}
  \label{2dPolar}
\end{equation}
where \ltheta, \lphi\ and \lthetaphi\ are the \jpsi\ polarization parameters. $\uptheta$ and $\upphi$ are the polar and azimuthal angles of the positively charged daughter lepton in the \jpsi\ rest frame with respect to a chosen quantization axis. In the helicity (HX) frame \cite{Jacob:1959at}, one uses the opposite of the direction of motion of the interaction point in the \jpsi\ rest frame as the quantization axis. In the Collins-Soper (CS) frame \cite{Collins:1977iv}, one chooses the bisector of the angle formed by one beam direction and the opposite direction of the other beam in the \jpsi\ rest frame. \jpsi\ is considered fully transversely or longitudinally polarized when the polarization parameters take the values of \mbox{(\ltheta, \lphi, \lthetaphi) = (1, 0, 0) or (-1, 0, 0)}. No polarization is referred to the case of (0, 0, 0). While the measured polarization values depend on the selection of the quantization axis, one can construct a frame invariant quantity to check the consistency of measurements in different frames \cite{Faccioli:2010kd}. It is defined as: 
\begin{equation}
\linv = \frac{\ltheta+3\lphi}{1-\lphi}.
\label{eq:inv}
\end{equation} 
Previous measurements of inclusive \jpsi\ polarization in 200 GeV \pp\ collisions \cite{Adare:2009js,Adamczyk:2013vjy} have only focused on \ltheta\ in the HX frame within $\pT<6$ \gev. In this paper, we extend the scope by measuring all three polarization parameters in both HX and CS frames for $\pT<10$ \gev, as well as the frame invariant quantity \linv. Measurements are carried out based on both the dimuon and dielectron decay channels covering different kinematic ranges. The inclusive \jpsi\ sample used in this paper includes directly produced \jpsi's and those from decays of excited charmonium states such as $\chi_{c}$ and $\psi$(2S) ($\sim$40\% \cite{Digal:2001ue}) as well as $b$-hadrons ($\sim$10-25\% above \pT \ of 5 \gev\ \cite{Adamczyk:2012ey}). These measurements will provide more stringent tests of different model calculations, especially for the universality of model parameters, such as $F_{\jpsi}$, LDMEs, that give models their predictive power. 

This paper is arranged as the following. An introduction to the Solenoidal Tracker At RHIC (STAR) is given in section \ref{sect:star}, followed by detailed descriptions of the analyses utilizing the electron and muon decay channels in sections \ref{sect:JpsiEE} and \ref{sect:JpsiMuMu}, respectively. The \jpsi\ polarization results are presented in section \ref{sect:result}, and a summary is given in section \ref{sect:summary}.

\section{STAR experiment}
\label{sect:star}
The STAR experiment \cite{Ackermann:2002ad} at RHIC consists of a suite of mid-rapidity detectors with excellent tracking and particle identification (PID) capabilities. The Time Projection Chamber (TPC) \cite{Anderson:2003ur} is a gaseous drift chamber with the readout system based on the Multi-Wire Proportional Chambers (MWPC) technology. It is the main tracking device to measure a particle's momentum and specific energy loss ($dE/dx$) for particle identification, and covers the pseudo-rapidity range of $|\eta| < 1$ over full azimuthal angle. A room temperature solenoidal magnet generates a uniform magnetic field of maximum value 0.5 T \cite{Bergsma:2002ac}. The Barrel Electromagnetic Calorimeter (BEMC) \cite{Beddo:2002zx} is a sampling calorimeter using lead and plastic scintillator. It is used to identify and trigger on high-\pT\ electrons over full azimuthal angle within $|\eta| < 1$. In conjunction with the start time provided by the Vertex Position Detector (VPD), the Time-Of-Flight (TOF) detector \cite{Bonner:2003bv} measures a particle's flight time to further improve the electron purity. For the muon channel analysis, the Muon Telescope Detector (MTD) \cite{Ruan:2009ug} is used for triggering on and identifying muons. It resides outside of the magnet which acts as an absorber, and covers about 45\% in azimuth within $|\eta| < 0.5$. Both the TOF and the MTD utilize the \mbox{Multi-gap} Resistive Plate Chamber (MRPC) technology. Forward-rapidity trigger detectors, the VPD at $4.24 < |\eta| < 5.1$ \cite{Llope:2014nva} and Beam-Beam Counters (BBC) at $3.3 < |\eta| < 5.0$ \cite{Kiryluk:2003aw}, are used to select collisions. 

\boldmath
\section{$\jpsi\rightarrow e^{+}e^{-}$}
\label{sect:JpsiEE}
\unboldmath

\subsection{Dataset, event and track selections}
\label{sect:EEData}
The dataset was taken for \pp\ collisions at \sqrts\ $= \mbox{200 GeV}$ in 2012 using both the  minimum-bias (MB) and high-tower (HT) triggers. The prescaled MB trigger selects non-single diffractive \pp\ collisions with a coincidence signal from the VPD on east and west sides, while the HT trigger selects events with energy depositions in the BEMC above given thresholds. About 300 million MB events, corresponding to an integrated luminosity of about $\mbox{~10 nb}$$^{-1}$, are analyzed to study the \jpsi\ polarization below \pT \ of 2 \gev. Data collected by the HT0 (HT2) trigger with an energy threshold of \mbox{$E_{\rm{T}} > 2.6$ (4.2) GeV} correspond to an integrated luminosity of 1.36 (23.5) $\text{pb}^{-1}$. The HT0 trigger is used for the \jpsi\ measurement within $2<\pT<4$ \gev, while the HT2 trigger is used for $4<\pT<14$ \gev.

The vertex position along the beam direction can be reconstructed from TPC tracks ($V_z^{\rm{TPC}}$) or from the time difference of east and west VPD signals ($V_z^{\rm{VPD}}$). A cut of $|V_{z}^{\rm{TPC}}|<50$ cm is applied to ensure good TPC  acceptance for all the events. An additional cut of $|V_z^{\rm{TPC}} - V_z^{\rm{VPD}}|<6$ cm is applied to reduce the pile-up background from out-of-time collisions for MB events.

Charged tracks are required to have at least 20 TPC space points (out of a maximum of 45), a ratio of at least 0.52 between actually used and maximum possible number of TPC space points, at least 11 TPC space points for $dE/dx$ calculation, and their distance of closest approach to the primary vertex (DCA) less than \mbox{1 cm}. Electrons and positrons are identified using $dE/dx$ in TPC, the velocity ($\beta$) calculated from the path length and time of flight between the collision vertex and TOF, and the ratio between the track momentum and energy deposition in the BEMC ($pc/E$)\cite{Luo_phdthesis}. The normalized $dE/dx$ is quantified as:
\begin{equation}
\nsigmae=\frac{\ln(dE/dx)_{\rm{measured}}-\ln(dE/dx)_{\rm{theory}}^{e}}{\sigma(\ln(dE/dx))},
\label{eq:nSigmaE}
\end{equation}
where $(dE/dx)_{\rm{measured}}$ is the measured energy loss in the TPC, $(dE/dx)_{\rm{theory}}^{e}$ is the expected energy loss for an electron based on the Bichsel formalism \cite{Bichsel:2006cs}, and $\sigma(\ln(dE/dx))$ is the resolution of the $\ln(dE/dx)$ measurement. The value of \nsigmae\ is required to be within (-1.9, 3). A cut of $|1/\beta - 1|<0.03$ is applied for TOF-associated candidates, and $0.3<pc/E<1.5$ is applied for BEMC-associated candidates above \mbox{1 \gev}. The electron and positron candidates are required to pass the \nsigmae\ cut, and either the $\beta$ or $pc/E$ cut. For HT-triggered events, at least one daughter of a \jpsi\ candidate must pass the $pc/E$ requirement and have an energy deposition in the BEMC higher than the HT trigger threshold.

\subsection{Analysis procedure}
\label{sect:EEAna}
A maximum likelihood method is used to extract all three \jpsi\ polarization parameters simultaneously. The likelihood is defined as:
\begin{widetext}
\begin{equation}
	-\ln{L(\ltheta, \lphi, \lthetaphi)} = -\sum N_{\jpsi} (\cos{\uptheta},\upphi) \ln{\left[F(\cos{\uptheta},\upphi \,|\, \ltheta, \lphi, \lthetaphi) \times \DetAccEff(\cos{\uptheta}, \upphi)\right]},
\label{eq:likelihood}
\end{equation}
\end{widetext}
where the sum is taken over the $(\cos{\uptheta}, \upphi)$ bins, $N_{\jpsi}(\cos{\uptheta},\upphi)$ is the raw number of \jpsi\ candidates  in each ($\cos{\uptheta}$, $\upphi$) bin, and $\DetAccEff(\cos{\uptheta}, \upphi)$ is the detector acceptance times \jpsi\ reconstruction efficiency in the same bin. $F(\cos{\uptheta},\upphi | \ltheta, \lphi, \lthetaphi)$ is the integral probability corresponding to $\cos{\uptheta}$ and $\upphi$ bin of positrons for given (\ltheta, \lphi, \lthetaphi) values, described by Eq.~\ref{2dPolar} normalized to 1. 
$\DetAccEff(\cos{\uptheta}, \upphi)$ is evaluated by simulating $\jpsi \rightarrow e^+ e^-$ decays, passing them through GEANT3 simulation \cite{Brun:1987ma} of the STAR detector, embedding the simulated digital signals into real data, and finally reconstructing the embedded events through the same procedure as for the real data. The central values and statistical errors of the \jpsi\ polarization parameters are obtained by maximizing the likelihood and corrected for possible biases that are estimated from a toy Monte Carlo (ToyMC). In this ToyMC, the same numbers of \jpsi\ signal and background candidates as in real data are randomly generated with fixed values of polarization parameters after applying detector acceptance and reconstruction efficiencies. The extracted \jpsi\ polarization parameters from the pseudo-data following the same procedure as described above are compared to the input values in terms of both central values and statistical errors, and the differences are applied as corrections to real data, which are generally very small compared to statistical errors. 

\subsection{Signal extraction}
\label{sect:EESig}
\begin{figure*}
\includegraphics[scale=0.9]{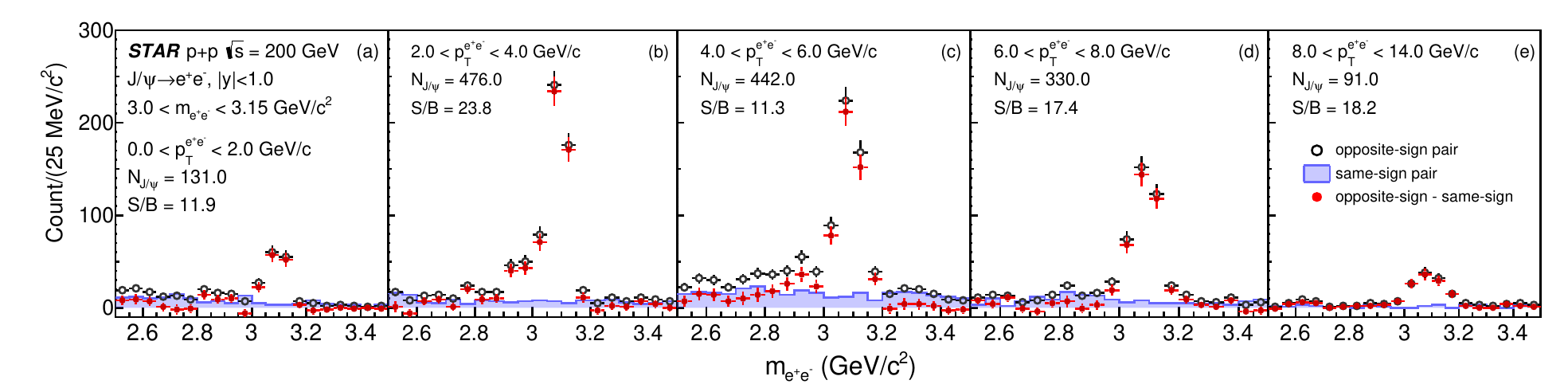}
\caption{\label{fig:jpsisignal}Invariant mass spectra of electron-positron pairs in different \pT\ bins (from left to right: \pT\ = 0-2, 2-4, 4-6, 6-8, 8-14 \gev). The black markers (blue filled histograms) are the spectra from opposite-sign (same-sign) charge pairs, while the red markers represent those obtained by subtracting the same-sign spectra from the opposite-sign ones. The number of \jpsi\ candidates ($N_{\jpsi}$) is given by the number of opposite-sign charge pairs minus that of same-sign charge pairs within \mbox{$3.0<m_{e^+e^-}<3.15$ \gevtwo}. The S/B ratio is the ratio between $N_{\jpsi}$ and that of the same-sign charge pairs within \mbox{$3.0<m_{e^+e^-}<3.15$ \gevtwo.}}
\end{figure*}

\begin{figure*}
\includegraphics[scale=0.9]{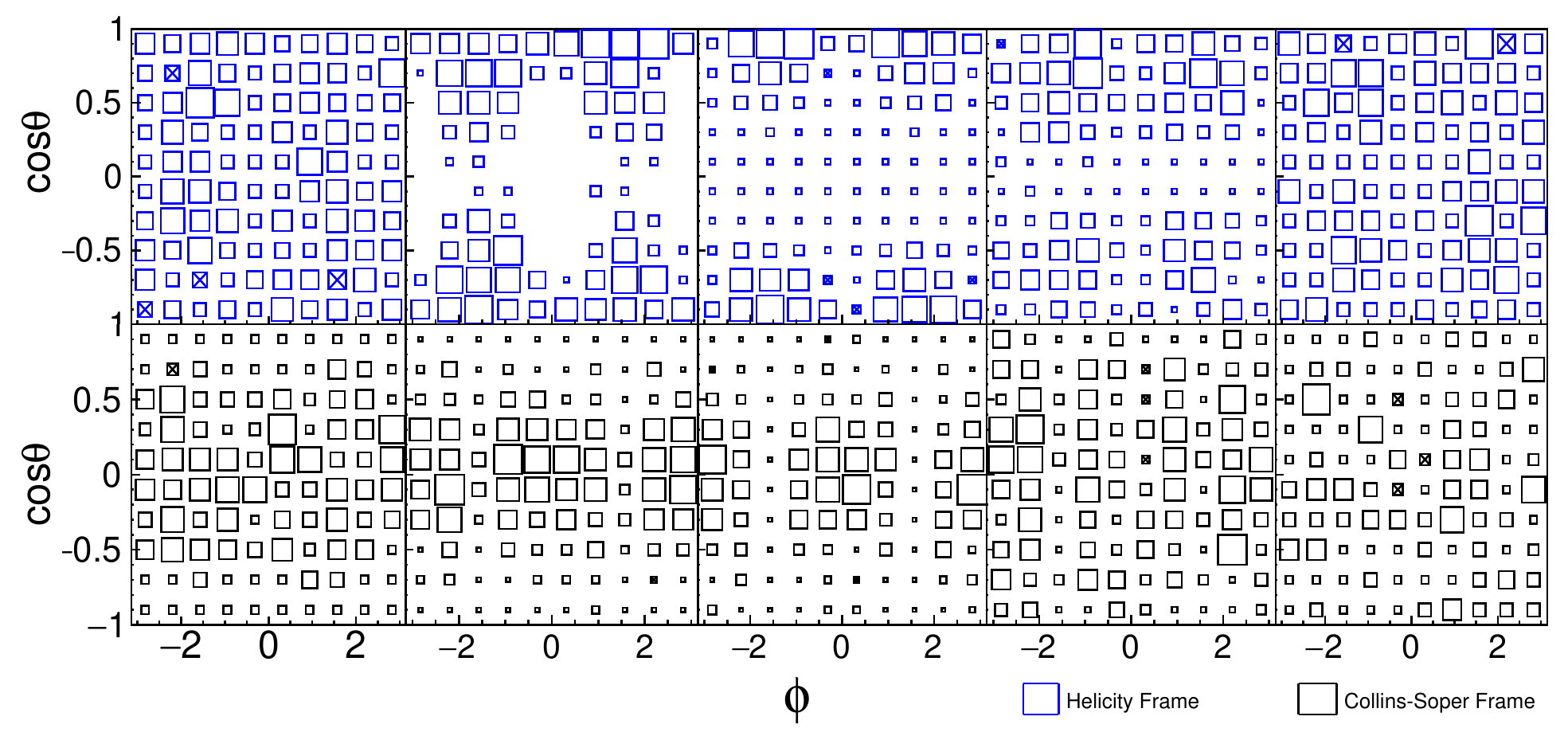}
\caption{\label{fig:data2d}Normalized two-dimensional $N_{\jpsi}(\cos{\uptheta},\upphi)/N^{total}_{\jpsi}$ distributions in different \pT\ bins (from left to right: \pT\ = 0-2, 2-4, 4-6, 6-8, 8-14 \gev). The top (bottom) row shows the distributions in the HX (CS) frame. The size of the boxes represents the absolute value of the \jpsi\ yield. The boxes with crosses are entries with negative values.}
\end{figure*}

\begin{figure*}
\includegraphics[scale=0.87]{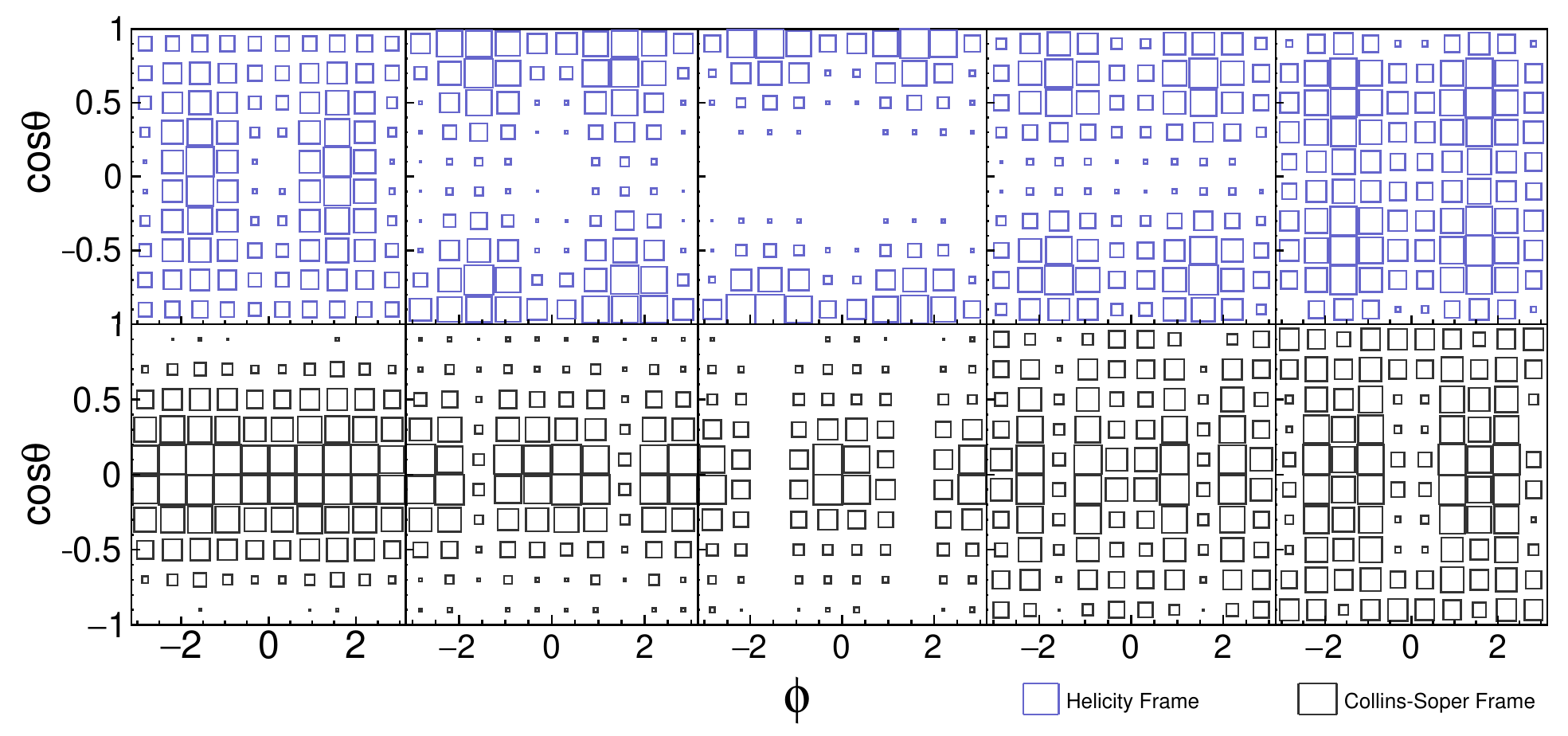}
\caption{\label{fig:eff2d}Two-dimensional $\DetAccEff(\cos{\uptheta}, \upphi)$ distributions in different \pT\ bins (from left to right: \pT\ = 0-2, 2-4, 4-6, 6-8, 8-14 \gev). The top (bottom) row shows the distributions in the HX (CS) frame. The size of the boxes represents the value of the \jpsi\ reconstruction efficiency times acceptance.}
\label{efficiency}
\end{figure*}

Invariant mass spectra of electron-positron pairs are shown in Fig.~\ref{fig:jpsisignal} for five different $\pT^{\jpsi}$ bins. The combinatorial background contribution is estimated by summing up the same-sign charge pairs of electron candidates ($e^-e^-$) and those of positron candidates  ($e^+e^+$), shown as filled areas in the figure. The raw numbers of \jpsi\ candidates are estimated by subtracting same-sign distributions from opposite-sign ones and integrating resulting counts within the invariant mass window of $3-3.15$ \gevtwo. The contribution from the residual background is found to be between 1.5-2.5\% and thus neglected here. The normalized two-dimensional $N_{\jpsi}(\cos{\uptheta},\upphi)/N_{\jpsi}^{total}$ distributions in the HX and CS frames are shown in Fig.~\ref{fig:data2d}.  The \jpsi\ reconstruction efficiency multiplied by the detector acceptance, $\DetAccEff(\cos{\uptheta}, \upphi)$, are shown in Fig.~\ref{fig:eff2d}, corresponding to the invariant mass window of $3-3.15$ \gevtwo. The detector acceptance, track reconstruction, BEMC electron identification and HT trigger efficiencies are estimated from simulation. Polarization of input \jpsi's does not play any role due to two-dimensional determination of the efficiencies. The electron identification efficiencies due to application of TPC and TOF requirements are estimated from data \cite{Adam:2018jmp} using a pure electron sample from gamma conversions. 
The electron $dE/dx$ and $1/\beta$ distributions are fit with a Gaussian distribution to calculate the cut efficiencies. The TOF matching efficiency is evaluated using TPC tracks that are matched to BEMC hits in order to suppress the pileup contribution. The bias due to the geometrical correlation between BEMC and TOF acceptance is corrected using an electron sample from data.

\begin{figure*}
\includegraphics[scale=0.87]{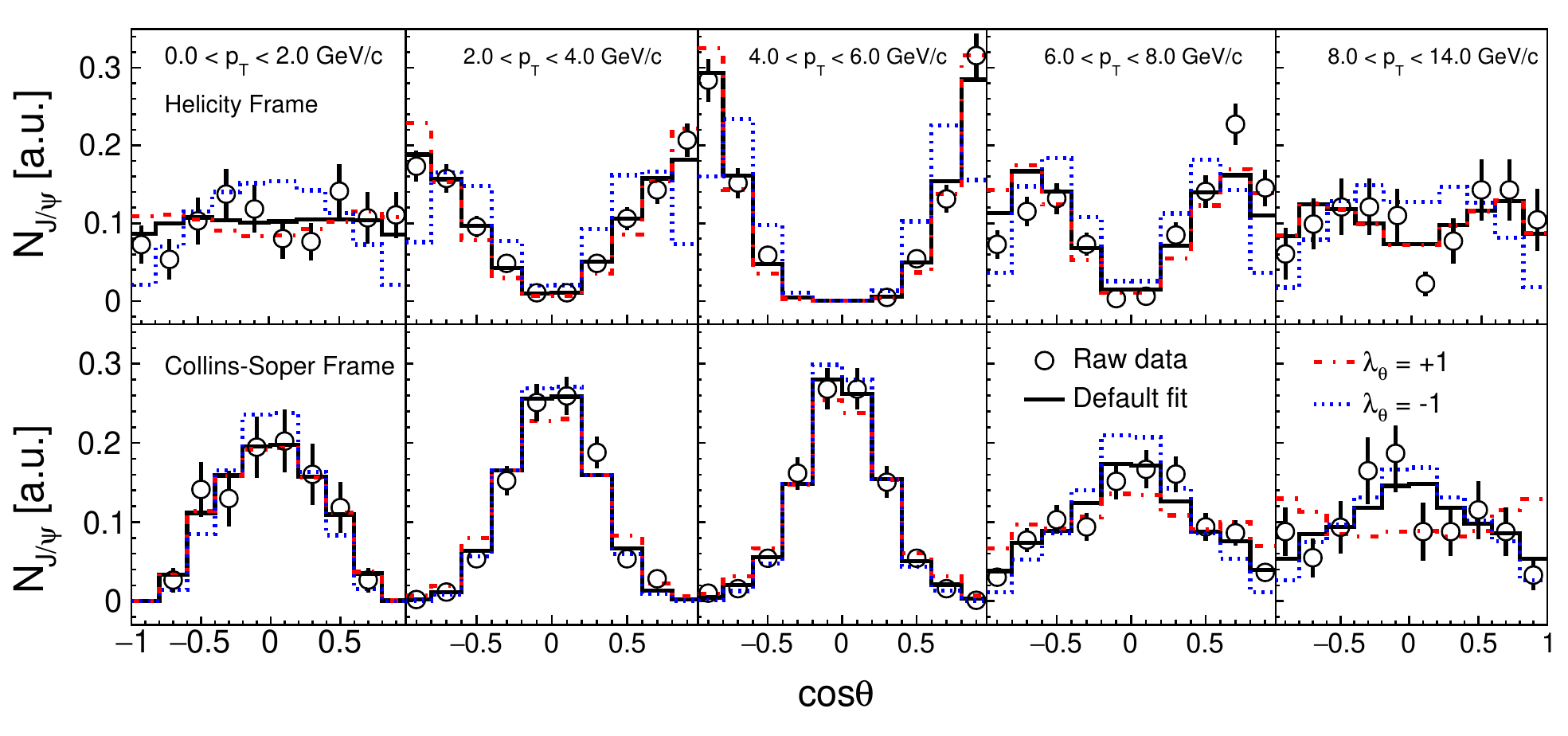}
\caption{The number of \jpsi\ candidates as a function of $\cos{\uptheta}$ in different \pT\ bins (from left to right: \pT\ = 0-2, 2-4, 4-6, 6-8, 8-14 \gev). The top (bottom) row shows the distributions in the HX (CS) frame. The solid lines correspond to the expected distributions based on the \jpsi\ polarization parameter values extracted from data. The dashed lines are the expected distributions with assumed values of $(\ltheta, \lphi, \lthetaphi) = (\pm1, 0, 0)$. The counts are after arbitrary normalization.}
\label{fig:costheta_comparison}

\includegraphics[scale=0.9]{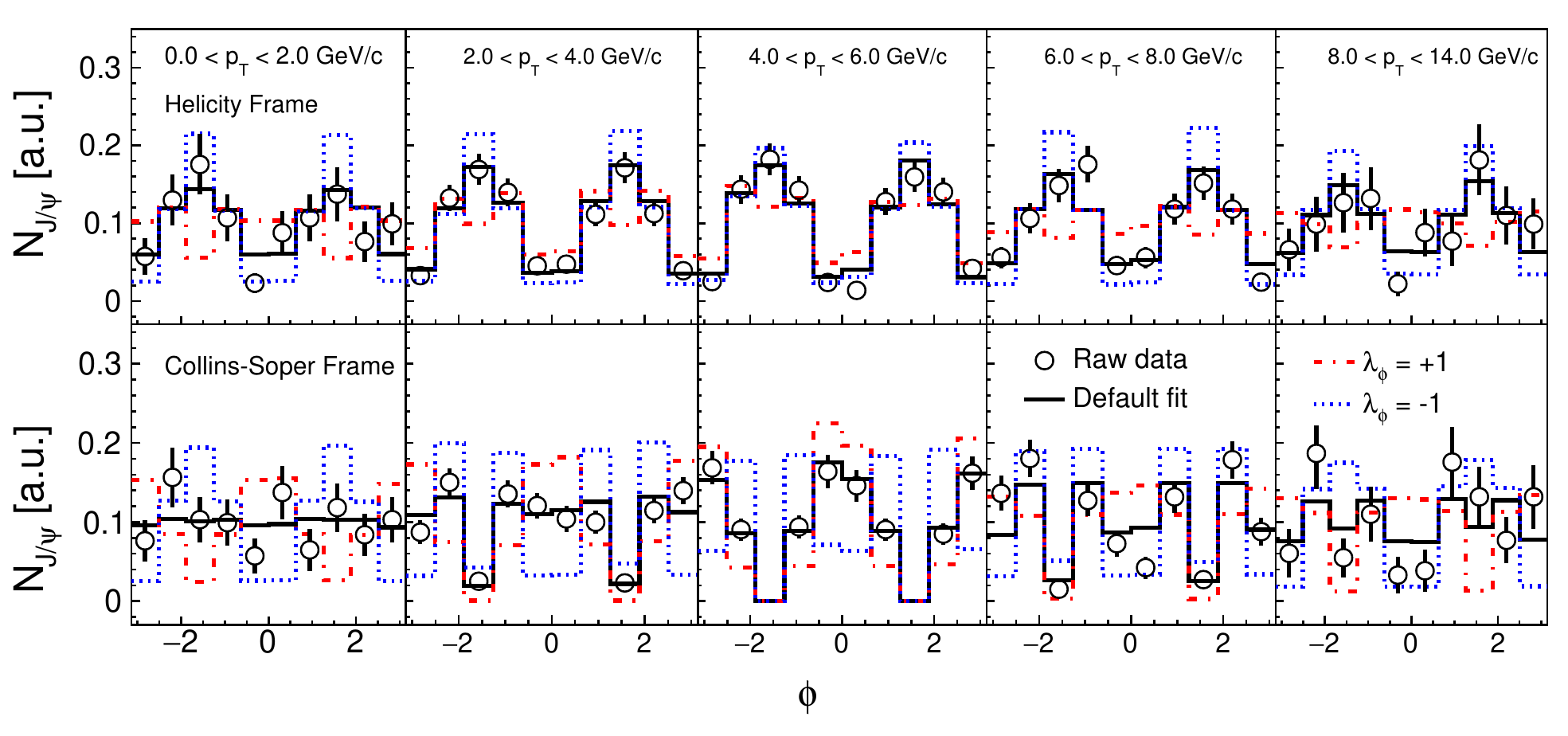}
\caption{The number of \jpsi\ candidates as a function of $\upphi$ in different \pT\ bins (from left to right: \pT\ = 0-2, 2-4, 4-6, 6-8, 8-14 \gev). The top (bottom) row shows the distributions in the HX (CS) frame. The solid lines correspond to the expected distributions based on the \jpsi\ polarization parameter values extracted from data. The dashed lines are the expected distributions with assumed values of $(\ltheta, \lphi, \lthetaphi) = (0, \pm1, 0)$. The counts are after arbitrary normalization.}
\label{fig:phi_comparison}
\end{figure*}

To check the results obtained from fit, the uncorrected \jpsi\ distributions are compared to the expected ones as shown in Figs.~\ref{fig:costheta_comparison} and \ref{fig:phi_comparison}. The former are obtained by projecting two-dimensional $N_{\jpsi}(\cos{\uptheta},\upphi)$ distributions onto either the $\cos{\uptheta}$ or $\upphi$ direction, while the latter are generated using the extracted \jpsi\ polarization parameters from data and taking into account the detector acceptance and efficiency. The expected \jpsi\ distributions agree well with the measured ones, confirming that the maximum likelihood method can be used to reliably extract the \jpsi\ polarization parameters. Also shown in these figures as references are the expected $\cos{\uptheta}$ and $\upphi$ distributions corresponding to the extreme cases where the polarization parameters $(\ltheta, \lphi, \lthetaphi) = (\pm1, 0, 0)$ and $(\ltheta, \lphi, \lthetaphi) = (0, \pm1, 0)$ are used, respectively.

The systematic uncertainties are estimated for the following sources. 
\begin{enumerate}[label=\arabic*)]
\item Acceptance: in extracting efficiencies from simulation, different parameterizations of the inclusive \jpsi\ \pT\ and rapidity spectra \cite{Adam:2018jmp} are tried and the difference is used as the uncertainty. 
\item PID: the uncertainty in the electron identification efficiencies is assessed by varying the mean and width of the TPC n$\sigma_{e}$ and TOF 1/$\beta$ distributions according to their uncertainties in the efficiency calculation, and by simultaneously varying the cut in both data and simulation on the ratio between the track momentum and energy deposition in the BEMC from $0.3<pc/E<1.5$ to $0.2<pc/E<1.4$ or $0.4<pc/E<1.6$. Additional uncertainties are considered in evaluating the TOF matching efficiency, including a correction factor to account for the correlation between BEMC and TOF acceptances in obtaining the TOF matching efficiency from data. 
\item Tracking: the uncertainty in track reconstruction efficiency is obtained by simultaneously varying the cuts in data and simulation on the minimum number of TPC hit points from 20 to 18, 19, 22, or 25, on maximum DCA from 1 cm to 0.8 or 1.2 cm, and by varying the momentum resolution in simulation within its uncertainty. 
\item Triggering: the uncertainty in the HT trigger efficiency is obtained by simultaneously changing the HT trigger threshold cut with $\pm$5\% variation. 
\end{enumerate}
For each of the systematic sources, the same analysis procedure is followed and the resulting maximum differences to the default results are taken as the uncertainties. 
The total systematic uncertainties are a quadrature sum of individual sources, as shown in Table \ref{tab:sys_dielectron}.

\begin{table}[htbp]
\renewcommand\arraystretch{1.3} 
\centering
\caption{Summary of systematic uncertainties for $\mbox{$\jpsi \rightarrow e^{+}e^{-}$}$ measurement.}
\begin{tabular}{ |p{1.6cm}<{\centering}|c|c|c|c|c|c|c| }
\hline  
 Source & \pT \, (\gev)& $\uplambda_{\uptheta}^{HX}$ & $\uplambda_{\upphi}^{HX}$ & $\uplambda_{\uptheta\upphi}^{HX}$ & $\uplambda_{\uptheta}^{CS}$ & $\uplambda_{\upphi}^{CS}$ & $\uplambda_{\uptheta\upphi}^{CS}$ \\[3pt] \hline 
\multirow{5}{*}{\shortstack{Acceptance}} 
&0-2    &   0.04 & 0.01 & 0.04 & 0.02 & 0.00 & 0.00 \\ 
&2-4    &   0.01 & 0.01 & 0.00 & 0.15 & 0.01 & 0.02 \\
&4-6    &   0.02 & 0.01 & 0.00 & 0.02 & 0.01 & 0.01 \\
&6-8    &   0.01 & 0.01 & 0.00 & 0.02 & 0.01 & 0.00 \\ 
&8-14  &   0.03 & 0.01 & 0.00 & 0.01 & 0.00 & 0.00 \\   \hline
\multirow{5}{*}{PID} 
&0-2    &   0.13 & 0.05 & 0.10 & 0.23 & 0.05 & 0.05 \\ 
&2-4    &   0.12 & 0.06 & 0.02 & 0.03 & 0.07 & 0.01 \\ 
&4-6    &   0.22 & 0.10 & 0.01 & 0.11 & 0.11 & 0.03 \\
&6-8    &   0.16 & 0.05 & 0.05 & 0.08 & 0.09 & 0.03 \\ 
&8-14  &   0.11 & 0.06 & 0.19 & 0.03 & 0.06 & 0.10 \\    \hline
\multirow{5}{*}{Tracking} 
&0-2    &   0.13 & 0.08 & 0.07 & 0.39 & 0.06 & 0.11 \\
&2-4    &   0.06 & 0.05 & 0.04 & 0.15 & 0.04 & 0.04 \\ 
&4-6    &   0.08 & 0.09 & 0.05 & 0.14 & 0.04 & 0.04 \\
&6-8    &   0.24 & 0.04 & 0.05 & 0.12 & 0.15 & 0.04 \\  
&8-14  &   0.33 & 0.05 & 0.12 & 0.06 & 0.10 & 0.11 \\   \hline
\multirow{5}{*}{Trigger} 
&0-2    &   0.00 & 0.00 & 0.00 & 0.00 & 0.00 & 0.00 \\
&2-4    &   0.04 & 0.03 & 0.01 & 0.01 & 0.02 & 0.02 \\ 
&4-6    &   0.17 & 0.17 & 0.01 & 0.14 & 0.02 & 0.10 \\ 
&6-8    &   0.10 & 0.03 & 0.03 & 0.06 & 0.03 & 0.03 \\
&8-14  &   0.12 & 0.00 & 0.01 & 0.01 & 0.03 & 0.00 \\  \hline
\multirow{5}{*}{Total} 
&0-2    &   0.19 & 0.10 & 0.12 & 0.45 & 0.08 & 0.12 \\
&2-4    &   0.14 & 0.09 & 0.05 & 0.21 & 0.08 & 0.05 \\ 
&4-6    &   0.29 & 0.22 & 0.05 & 0.22 & 0.12 & 0.11 \\
&6-8    &   0.30 & 0.07 & 0.07 & 0.15 & 0.18 & 0.06 \\ 
&8-14  &   0.37 & 0.08 & 0.22 & 0.07 & 0.12 & 0.15 \\  \hline
\end{tabular}
\label{tab:sys_dielectron}
\end{table}

\boldmath
\section{$\jpsi\rightarrow\mu^{+}\mu^{-}$}
\label{sect:JpsiMuMu}
\unboldmath

\subsection{Dataset, event and track selections}
\label{sect:MuMuData}
The dataset was taken for \pp\ collisions at \mbox{\sqrts\ = 200 GeV} in 2015, and corresponds to an integrated luminosity of 122 \invpb. Events are selected online with a dimuon trigger, which requires at least two signals in the MTD whose timing difference to the start time provided by the VPD falls within the pre-defined trigger timing window. 

Events used in offline analysis are required to have a vertex position of $|V_{z}^{\rm{TPC}}|<100$ cm along the beam direction to maximize statistics. 
Primary vertices are further required to be within 2 cm radially with respect to the center of the beam pipe. 

In the analysis of the dimuon decay channel, charged tracks reconstructed in the TPC should have at least 15 TPC space points used for reconstruction. The ratio of the actually used to the maximum possible number of TPC space points is required to be larger than 0.52 to reject split tracks. The distance of closest approach (DCA) to the primary vertex needs to be less than 3 cm to suppress contribution from secondary decays and pile-up tracks. The selected TPC tracks are afterwards refit with the primary vertex included in order to improve the momentum resolution. Tracks are then propagated from the TPC to the MTD radius. Only tracks with $\pT> 1.3$ \gev\ are selected to achieve high efficiency for reaching the MTD after losing energy along the trajectory. Once a track is matched to the closest MTD hit, cuts on variables, \deltay, \deltaz\ and \deltatof, are applied to further suppress background hadrons. Here, \deltay\ and \deltaz\ are the residuals between the projected track position at the MTD radius and the matched MTD hit along azimuthal and beam directions, respectively. We require \deltay\ and \deltaz\ to be within 3 (3.5) $\sigma$ of their resolutions for \pT\ $<$ $(>)$ 3 \gev. \deltatof\  is the difference between the measured time-of-flight with the MTD and the calculated time-of-flight from track extrapolation with a muon particle hypothesis, and should satisfy $|\deltatof|<1$ ns. Additional PID capabilities arise from the energy loss measurement in the TPC. It is quantified as \nsigmapi, whose definition is similar to that of electrons as described in Sect. \ref{sect:EEData} but using a pion hypothesis. In the kinematic range relevant for this analysis, muons are expected to lose more energy than pions by about half of $dE/dx$ resolution. A cut of $-2 < \nsigmapi < 3$ is applied.

\subsection{Analysis procedure}
\label{sect:MuMuAna}
To extract the \jpsi\ polarization in the dimuon decay channel, a different strategy is adopted compared to the one used for the dielectron channel as described in Sect. \ref{sect:EEAna}. Equation \ref{2dPolar} is integrated over $\upphi$ and $\cos\uptheta$, yielding two 1-D distributions:
\begin{equation}
W(\cos\uptheta) \propto 1+\uplambda_{\uptheta} \cos^{2}\uptheta,
\label{eq:theta}
\end{equation}
and 
\begin{equation}
W(\upphi)\propto1+\frac{2\uplambda_{\upphi}}{3+\uplambda_{\uptheta}}\cos{2\upphi},
\label{eq:phi}
\end{equation}
The \lthetaphi\ term vanishes in both integrations. The polarization parameters, \ltheta\ and \lphi, are extracted from a simultaneous fit to corrected \jpsi\ yield distributions as a function of $\cos\uptheta$ and $\upphi$ of daughter $\mu^{+}$  with Eqs. \ref{eq:theta} and \ref{eq:phi}. This strategy is motivated by the worse signal-to-background ratio for the dimuon decay channel compared to the dielectron decay channel, and fitting the 1-D distributions of Eqs. \ref{eq:theta} and \ref{eq:phi} is therefore more stable. However, the $\uplambda_{\uptheta\upphi}$ parameter cannot be extracted from this method.

The number of \jpsi\ extracted in each $\cos\uptheta$ or $\upphi$ bin needs to be corrected for the detector acceptance and efficiency, denoted as \DetAccEff$(\cos\uptheta, \upphi)$. 
It is evaluated via simulation as described in Sect. \ref{sect:EEAna} but for the muon channel. Since \DetAccEff$(\cos\uptheta, \upphi)$ depends on both $\cos\uptheta$ and $\upphi$, the projected 1-dimentional(1-D) \DetAccEff\ as a function of $\cos\uptheta$ or $\upphi$ is affected by the assumed polarization of input \jpsi\ in the simulation. On the other hand, the \lthetaphi\ value does not affect the averaged \DetAccEff\ as \DetAccEff$(\cos\uptheta, \upphi)$ is symmetric with respect to $\cos\uptheta=0$ and $|\upphi-\pi/2|=0$. Given that the \jpsi\ polarization is not known {\it a priori} and the correction for the detector acceptance and efficiency depends on it, an iterative procedure is adopted. In the first iteration, the 1-D \DetAccEff\ as a function of $\cos\uptheta$ or $\upphi$ is evaluated using non-polarized \jpsi\ in the simulation, and the polarization parameters are extracted from data after correcting for \DetAccEff. In the second iteration, the extracted polarization parameters from the previous iteration are used in the simulation to assess \DetAccEff, which in turn is used to correct data and obtain new polarization parameters. The iteration continues until the differences of the obtained polarization parameters between two consecutive iterations are less than 0.01. This threshold is determined based on the statistical precision of the data. 

\begin{figure}[htbp]
\includegraphics[scale=0.43]{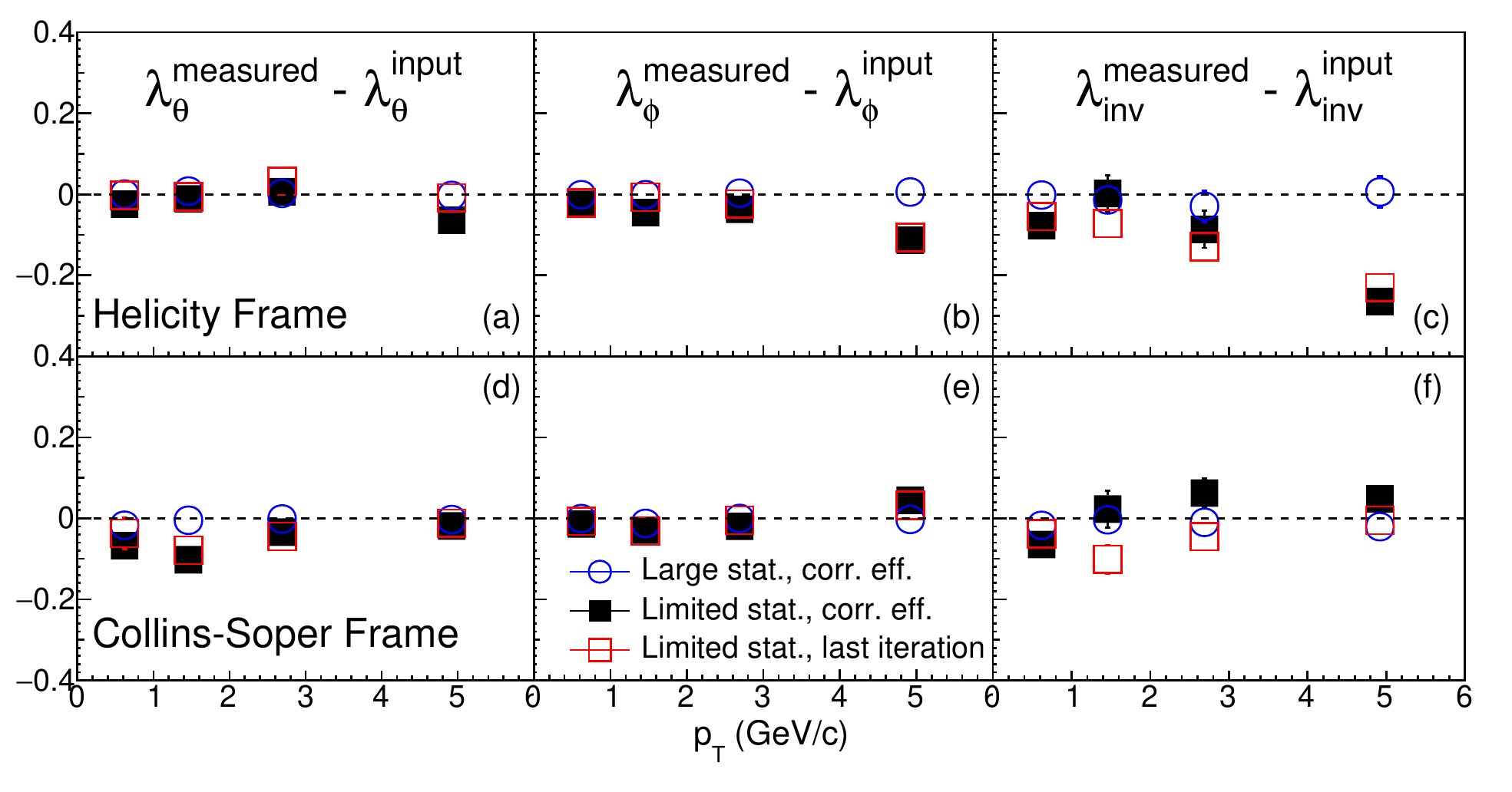}
\caption{Differences of the extracted \jpsi\ polarization parameters, \ltheta, \lphi\ and \linv, in the ToyMC and the input values in both the HX and CS frames. Data points of different markers correspond to the three different tests. See text for details.}
\label{fig:last_iter_toy}
\end{figure}

To validate the iterative procedure, a ToyMC is developed which is different from that used in the electron channel analysis. The single muon efficiency as a function of \pT, $\eta$ and $\upphi$, extracted from the GEANT simulations, is applied to mimic realistic detector acceptance and detection efficiency. \jpsi's with realistic \ltheta\ and \lphi\ values in four different \pT\ bins, as presented in Sec. \ref{sect:result}, are used as input to the ToyMC while the \lthetaphi\ value is assumed to be 0.  Both pseudo-data and \jpsi\ pseudo-efficiency are generated in the ToyMC. Depending on the statistical precision of the pseudo-data and how the pseudo-efficiency is obtained, the following tests are done: 
\begin{itemize}
\item Test 1 -- Large statistics with correct efficiency (``Large stat., corr. eff."): the pseudo-data sample has significantly larger statistics than real data, and the pseudo-efficiency is generated using the same polarization parameters as for pseudo-data. This represents a best-case scenario, and the polarization parameters, \ltheta, \lphi and \linv, are extracted using Eqs. \ref{eq:theta} and \ref{eq:phi}. Differences to the input polarization values are shown in Fig. \ref{fig:last_iter_toy} as open circles for both HX and CS frames. In most cases, the input values are recovered with small discrepancies arising from the limited acceptance of the MTD. 
\item Test 2 -- Limited statistics with correct efficiency (``Limited stat., corr. eff."): the pseudo-data sample has comparable statistical precision to real data, while the pseudo-efficiency is generated using the same polarization parameters as for pseudo-data. To avoid random fluctuation of one pseudo-data sample, 500 independent samples of similar statistics are generated. The mean values of the polarization parameters extracted from the 500 pseudo-data samples are compared to the input values, and the differences are shown in Fig. \ref{fig:last_iter_toy} as filled squares. Compared to ``Test 1", the extracted polarization values deviate further from the input ones due to the influence of the limited statistics in the pseudo-data sample on top of the limited MTD acceptance. The relatively large deviation seen in \linv\ around 5 \gev\ in the HX frame is an amplification of the smaller, but still sizable deviation seen in \lphi.
\item Test 3 -- Limited statistics using the iterative procedure (``Limited stat., last iteration"): in the last test, the 500 pseudo-data samples are generated with comparable statistical precision to real data, but the iterative procedure as described above is used to obtain the efficiency. Polarization values equal to 0 are used in the first iteration, and the procedure stops after the same convergence criterion of 0.01, as for the real data, between two consecutive iterations is fulfilled. 
The resulting differences to the input values are shown as open squares in Fig. \ref{fig:last_iter_toy}, and agree with ``Test 2" quite well. This indicates that very small biases are introduced in the iterative procedure. 
\end{itemize}
It has also been found that the correct polarization values are always obtained as long as the convergence occurs and no matter what input polarization values are used in the first iteration. The ToyMC validation confirms that the \jpsi\ polarization parameters can be extracted reliably using the iterative procedure. The residual biases shown in Fig. \ref{fig:last_iter_toy} are corrected for, as described in Sect. \ref{sect:MuMuSig}. 

\subsection{Signal extraction}
\label{sect:MuMuSig}
\begin{figure}[htbp]
\centering
\includegraphics[scale=0.4]{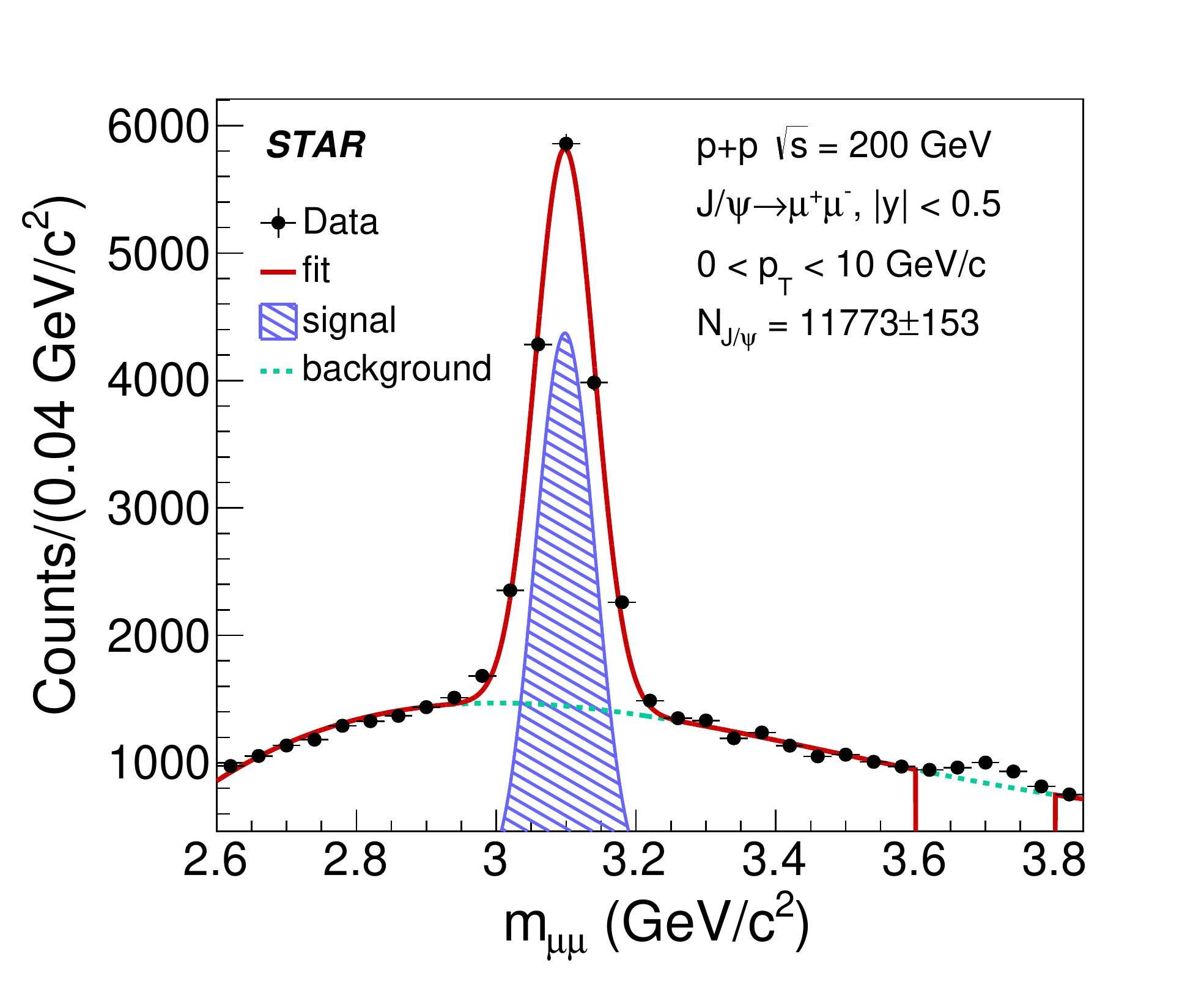}\centering
\caption{The invariant mass spectrum of opposite-sign muon pairs for $\pT<10$ \gev. The solid line depicts the fitting result, consisting of a Gaussian function representing the signal (shaded area) and a polynomial function describing the background (dashed line). Data points in the  $\psi$(2S) mass region ($3.6 < \text{M}_{\mu\mu} < 3.8$ \gevtwo) are excluded from the fit. }
\label{fig:signal}
\end{figure}

\begin{figure}[htbp]
\centering
\includegraphics[scale=0.43]{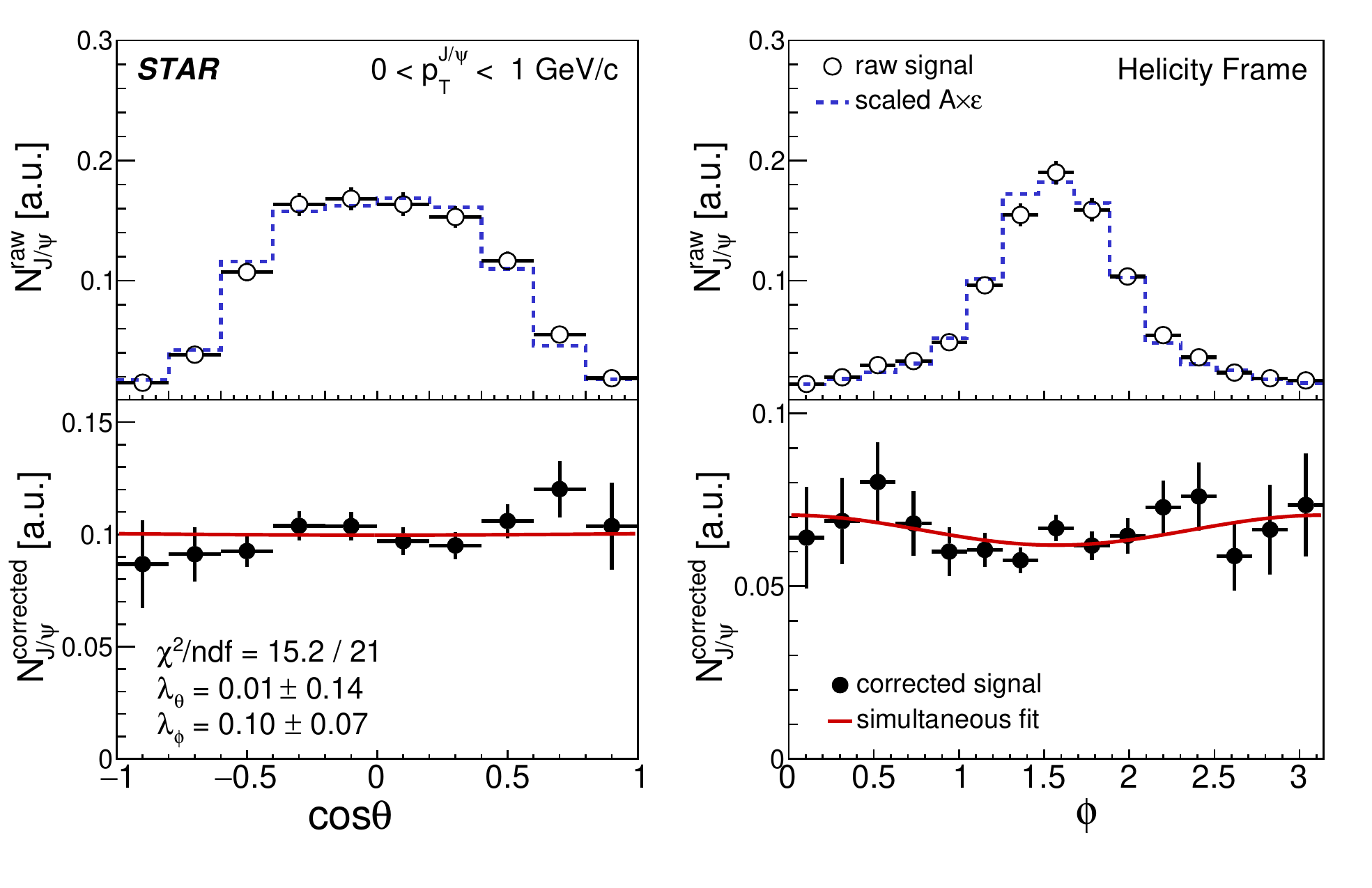}
\caption{Upper: the raw \jpsi\ yield and \DetAccEff\ from the last iteration as a function of $\mu^{+}$ $\cos\uptheta$  (left) and $\upphi$ (right) in the HX frame for $0<\pT^{\jpsi}<1$ \gev. Lower: the acceptance and efficiency corrected \jpsi\ yields along with the simultaneous fit. The counts are after arbitrary normalization.}
\label{fig:last_iter_HX}
\end{figure}

The selected muon candidates of opposite-sign charges are paired, and the resulting invariant mass distribution is shown in Fig. \ref{fig:signal} for the entire sample used in the dimuon channel analysis. 
The raw \jpsi\ yield is extracted by fitting the invariant mass distribution with a Gaussian function describing the \jpsi\ signal, and a polynomial function describing the background. Data points in the  $\psi$(2S) mass region ($3.6 < \text{M}_{\mu\mu} < 3.8$ \gevtwo) are excluded from the fit. The mean of the Gaussian distribution is fixed to the \jpsi\ mass in the PDG \cite{Tanabashi:2018oca}. In total, the \jpsi\ yields are extracted in ten $\cos\uptheta$ bins and fifteen $\upphi$ bins for each $\pT^{\jpsi}$ interval. The order of the background polynomial function ranges from 2 to 5, depending on the $\pT^{\jpsi}$ and the $\mu^{+}$ $\cos\uptheta$ and $\upphi$ bin. 
To facilitate the fits below 2 \gev, the widths of the Gaussian function in individual $\cos\uptheta$ and $\upphi$ bins are fixed to be the same as that extracted 
from fitting the inclusive invariant mass distribution integrated over $\cos\uptheta$ and $\upphi$ bins in the same $\pT^{\jpsi}$ interval. For $\pT^{\jpsi}$ above 2 \gev, the width of the Gaussian function is left as a free parameter.
Variations in the following aspects of the fit procedure are applied: 
the bin width of the invariant mass distribution, 
fixing the width of the Gaussian function also for $\pT^{\jpsi}$ above 2 \gev,  the order of the polynomial function and the fit range. The average \jpsi\ yields from these variations are used for extracting the polarization parameters. \jpsi\ yields with significance less than 3 are not considered. Upper panels of Fig. \ref{fig:last_iter_HX} show an example of the average raw \jpsi\ yield, depicted as open circles, as a function of $\cos\uptheta$ and $\upphi$ for \mbox{$0<\pT^{\jpsi}<1$ \gev} in the HX frame. 

\begin{figure}[htbp]
\centering
\includegraphics[scale=0.43]{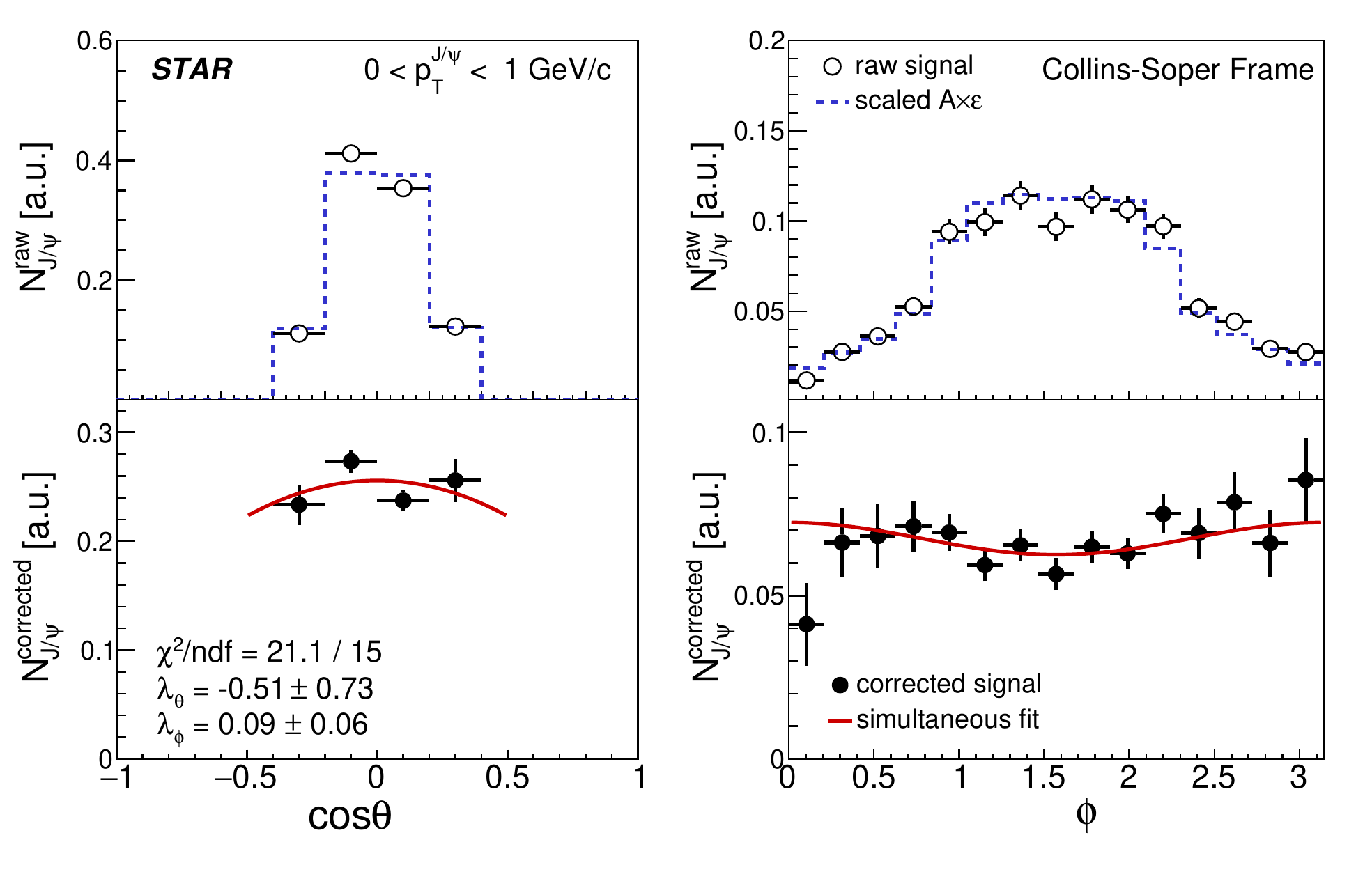}
\caption{Upper: the raw \jpsi\ yield and \DetAccEff\ from the last iteration as a function of $\mu^{+}$ $\cos\uptheta$  (left) and $\upphi$ (right) in the CS frame for $0<\pT^{\jpsi}<1$ \gev. Lower: the acceptance and efficiency corrected \jpsi\ yields along with the simultaneous fit. The counts are after arbitrary normalization.}
\label{fig:last_iter_CS}
\end{figure}

Following the iterative procedure, the efficiency multiplied by the detector acceptance from the last iteration is shown in the upper panel of Fig. \ref{fig:last_iter_HX} as dashed lines. It is scaled to the same integral as the data distribution. The lower panels of Fig. \ref{fig:last_iter_HX} show the fully corrected \jpsi\ yield as a function of $\cos\uptheta$ and $\upphi$, along with the simultaneous fit to both distributions using Eqs. \ref{eq:theta} and \ref{eq:phi} (solid lines). The polarization parameters, \ltheta\ and \lphi, are obtained from the simultaneous fit and listed in the figure. Similar plots in the CS frame are shown in Fig. \ref{fig:last_iter_CS} for \mbox{$0<\pT^{\jpsi}<1$ \gev}. The average $\uplambda_{\uptheta}$ and $\uplambda_{\upphi}$ values from 24 combinations of different track quality and muon identification cuts are taken as the central values. 
Figures \ref{fig:last_iter_HX} and \ref{fig:last_iter_CS} show results from one such combination as an example.
As shown in Fig. \ref{fig:last_iter_toy}, small biases are present in the extracted polarization parameters, due to a combination of limited MTD acceptance, limited statistical precision of data and the usage of the iterative procedure. Using the ToyMC described above, the extracted \jpsi\ polarization parameters are compared with input values in terms of central values and statistical errors, and the differences are applied as corrections to real data, which are also very small compared to statistical errors.

The systematic uncertainties arise from signal extraction, detector acceptance and efficiency. 
As mentioned above, different aspects of the fitting procedure are varied, and 24 combinations of track quality and muon identification cuts are tried. For the latter, cuts are changed consistently in data and MC simulation, and the entire analysis chain is repeated for each case. The RMS's of the distributions for polarization parameters stemming from these two groups of variations are taken as the systematic uncertainties.
The total systematic uncertainties are a quadrature sum of individual sources, as shown in Table \ref{tab:sys_dimuon}.

\begin{table}[H]
\renewcommand\arraystretch{1.3} 
\centering
\caption{Summary of systematic uncertainties for $\mbox{$\jpsi\rightarrow \mu^{+}\mu^{-}$}$ measurement.}
\label{tab:sys_dimuon}
\begin{tabular}{ |c|c|c|c|c|c| }
\hline
Source & \pT \, (\gev) & $\uplambda_{\uptheta}^{HX}$ & $\uplambda_{\upphi}^{HX}$ & $\uplambda_{\uptheta}^{CS}$ & $\uplambda_{\upphi}^{CS}$  \\ \hline
\multirow{4}{*}{\tabincell{c}{Signal \\ extraction}} 
&0-1   &  0.10 &  0.02  & 0.19 &  0.03  \\ 
&1-2    &  0.29 &  0.05  & 0.31 &  0.09 \\ 
&2-4    &  0.05 &  0.03  & 0.09 &  0.06 \\  
&4-10  &  0.15 &  0.05  & 0.09 &  0.20  \\  \hline
\multirow{4}{*}{\tabincell{c}{Tracking \\ and \\ PID}} 
&0-1    & 0.04 & 0.02  & 0.24 &  0.02  \\ 
&1-2    & 0.11 & 0.07  & 0.33 &  0.07  \\ 
&2-4    & 0.10 & 0.10  & 0.12 &  0.08 \\  
&4-10  & 0.06 & 0.05  & 0.12 &  0.08  \\  \hline
\multirow{4}{*}{Total} 
&0-1    &  0.11 &  0.03  & 0.32 & 0.04  \\ 
&1-2    &  0.31 &  0.09  & 0.45 &  0.11  \\ 
&2-4    &  0.11 &  0.10  & 0.16 &  0.10  \\  
&4-10  &  0.17 &  0.07  &0.15 &  0.22  \\  \hline
\end{tabular}
\end{table}

\section{Results}
\label{sect:result}
\begin{figure*}[ht]
\includegraphics[scale=0.7]{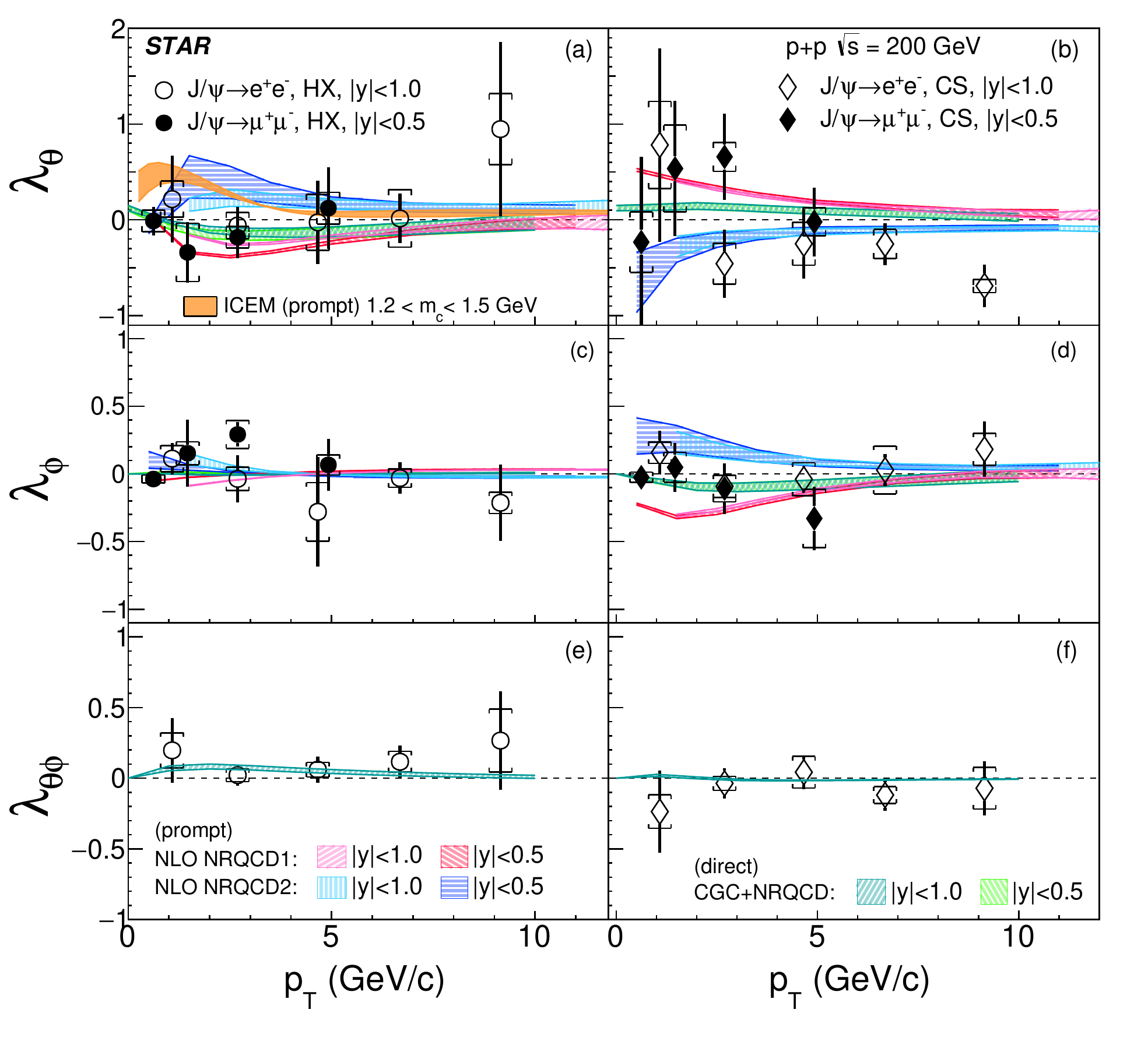}
\caption{The \jpsi\ polarization parameters (from top to bottom: \ltheta, \lphi, \lthetaphi) as a function of \pT\ in the HX (left) and CS (right) frames. Open and filled symbols are for measurements through the dielectron and dimuon decay channels covering different rapidity ranges. The vertical bars represent the statistical errors while the boxes around data points depict the systematic uncertainties. The data points are placed at the mean \pT\ value determined from the inclusive $\pT^{\jpsi}$ spectrum measured in \pp\ collisions at \sqrts\ = 200 GeV \cite{Adam:2018jmp}. Model calculations \cite{Cheung:2018tvq, Ma:2018qvc, Zhang:2014ybe, Gong:2012ug} are also shown for comparison. The ICEM and two NLO NRQCD calculations are for prompt \jpsi, while the CGC+NRQCD is for direct \jpsi.}
\label{fig:final_NRQCD}
\end{figure*}

The polarization parameters, \ltheta, \lphi\ and \lthetaphi\ for inclusive \jpsi\ are measured in both the HX and CS frames in \pp\ collisions at \sqrts\ = 200 \gev, as shown in Fig. \ref{fig:final_NRQCD}. The dimuon and dielectron results are shown as filled and open symbols respectively, and are consistent with each other in the overlapping \pT\ range even though they cover different rapidity regions. Measurements of $\lambda_{\theta\phi}$ via the dimuon channel, currently not available, will be carried out in the future with a larger data sample than the one used in this paper. All three polarization parameters are consistent with 0 within statistical and systematic uncertainties, except for \ltheta\ in the CS frame above 8 \gev\ whose central value is at \mbox{--0.69 $\pm$ 0.22 $\pm$ 0.07}. No strong \pT\ dependence is seen in all cases. The numerical values of the measured \jpsi\ polarization parameters are listed in Appendix (Tables \ref{Final_stas_HX_EE}$-$\ref{Final_stas_CS_MuMu}). Model calculations for prompt \jpsi\ from ICEM \cite{Cheung:2018tvq}, NRQCD with two sets of LDMEs denoted as ``NLO NRQCD1" \cite{Zhang:2014ybe} and ``NLO NRQCD2" \cite{Gong:2012ug}, are shown in Fig. \ref{fig:final_NRQCD} for comparison. 
Non-prompt \jpsi\ from b-hadron decays, not included in aforementioned model calculations, make about 10-25\% of the inclusive \jpsi\ sample above 5 GeV/c, with the fraction decreasing to be negligible at 1 \gev\ \cite{Chao:2012iv}. The effective polarization for non-prompt \jpsi\ above 5 \gev\ within $|y| <$ 0.6 is measured to be \mbox{\ltheta\ =  --0.106 $\pm$ 0.033 $\pm$ 0.007} in $p$+$\overline{p}$ collisions at $\sqrt{\text s}$ = 1.96 TeV \cite{CDF}. Therefore, contribution to the inclusive \jpsi\ polarization from b-hadron decays is expected to be small. Also shown in Fig. \ref{fig:final_NRQCD} are CGC+NRQCD calculations for direct \jpsi, in which both non-prompt \jpsi\ and those from decays of excited charmonium states are not included \cite{Ma:2018qvc}. A recent CMS measurement supports that at least one of $\chi_\mathrm{c1}$ and $\chi_\mathrm{c2}$ is strongly polarized in the HX frame in \pp\ collisions at 8 TeV, in agreement with NRQCD predictions \cite{chi_c}. It has been checked explicitly in [19] that feeddown corrections from $\chi_\mathrm{cJ}$ states on \jpsi\ polarization parameters are small and within theoretical uncertainties.
For \ltheta\ in the HX frame, the ICEM calculation predicts a sizable transverse polarization at low \pT, while the \jpsi\ polarization from CGC+NRQCD changes from slightly transverse at low \pT\ to slightly longitudinal at higher \pT. The difference between the LDMEs used in the two NLO NRQCD calculations is that additional $\eta_{c}$ production data measured by the LHCb Collaboration \cite{Aaij:2014bga} is used to determine LDMEs for ``NLO NRQCD1" besides those used for the case of ``NLO NRQCD2". They show opposite behaviors for \ltheta\ and \lphi\ in both reference frames. To quantify the agreement between data and model calculations, the $\chi^{2}$ test has been performed simultaneously using the data points in HX and CS frames for both channels. The $\chi^{2}$/NDF and corresponding $p$-values are listed in Table \ref{tb:chi2}.

\begin{table}[htbp]
\renewcommand\arraystretch{1.3} 
\centering
\caption{List of $\chi^{2}$/NDF and the corresponding $p$-values between data of inclusive \jpsi\ polarization and different model calculations of prompt or direct \jpsi\ polarization.}
\label{tb:chi2}
\begin{tabular}{ c|c|c}
\hline
Model & $\chi^{2}$/NDF & $p$-value \\ \hline
ICEM \cite{Cheung:2018tvq} & 13.28/9 & 0.150  \\ \hline
NRQCD1 \cite{Zhang:2014ybe} & 48.81/32 & 0.029  \\ \hline
NRQCD2 \cite{Gong:2012ug} & 42.99/32 & 0.093  \\ \hline
CGC+NRQCD \cite{Ma:2018qvc} & 32.11/46 & 0.940 \\ \hline
\end{tabular}
\end{table}
While no model can be ruled out definitively based solely on the data presented, the CGC+NRQCD gives the best overall description.

\begin{figure}[ht]
\includegraphics[width=0.5\textwidth]{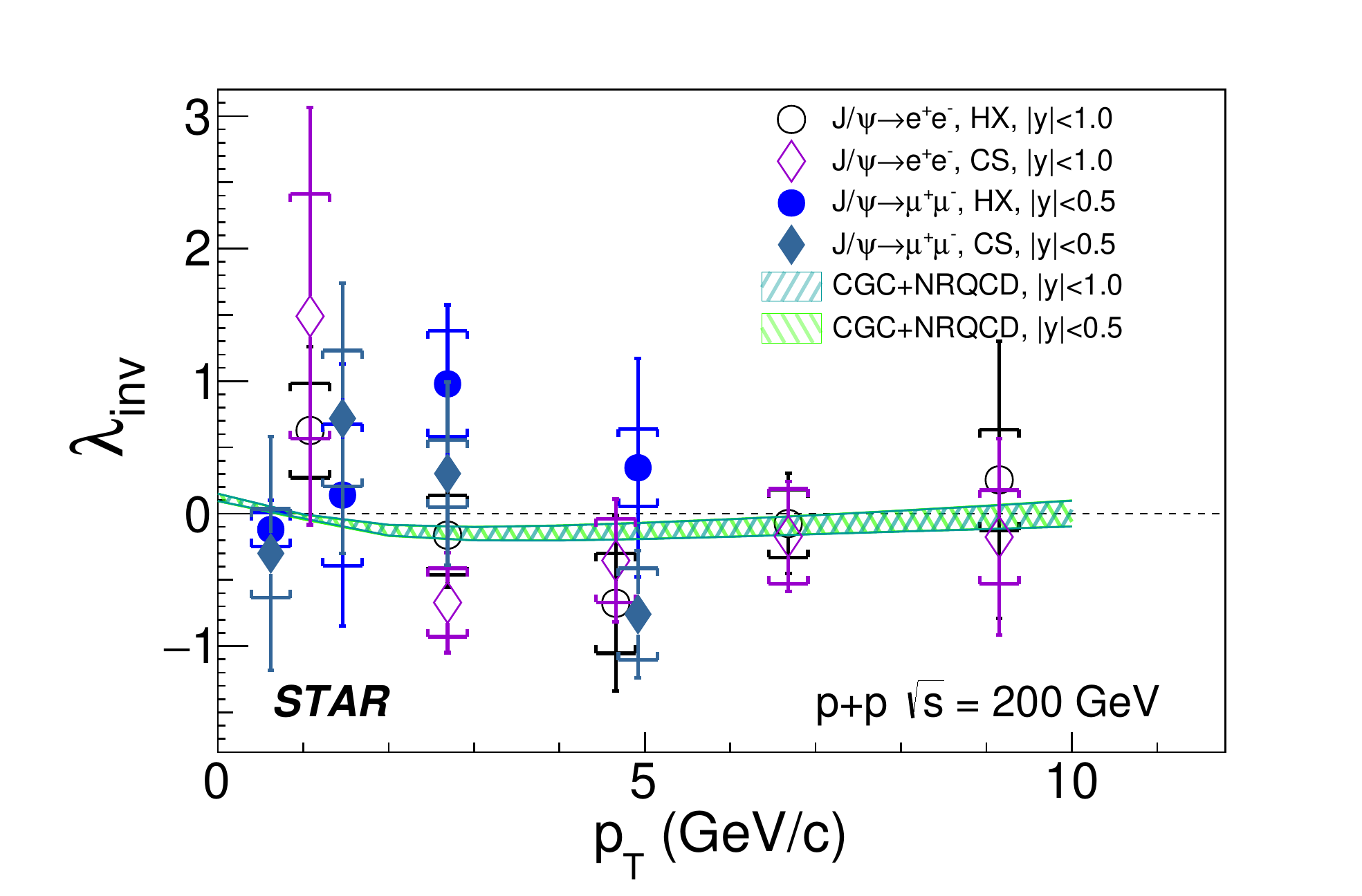}
\caption{\linv\ of \jpsi\ vs. \pT\ in both HX (circles) and CS (diamonds) reference frames. The open and filled symbols are for measurements through the dielectron and dimuon decay channels, respectively. The vertical bars represent the statistical errors while the boxes around data points depict the systematic uncertainties.  CGC+NRQCD \cite{Ma:2018qvc} calculations are also shown for comparison. }
\label{fig:lambda_inv}
\end{figure}

The \linv\ values extracted according to Eq. \ref{eq:inv} for inclusive \jpsi\ are shown in Fig. \ref{fig:lambda_inv}  as a function of \pT\ for both the HX and CS frames. The dimuon and dielectron results are shown as filled and open circles, respectively. The vertical bars represent the statistical errors while the boxes around data points depict the systematic uncertainties. The \linv\ values measured in the two frames are consistent with each other within experimental uncertainties, confirming the reliability of the results. The \linv\ values are consistent with the CGC+NRQCD calculations within uncertainties.   

\section{Summary}
\label{sect:summary}
For the first time, the inclusive \jpsi\ polarization parameters, \ltheta, \lphi\ and \lthetaphi, are measured as a function of \pT\ in \pp\ collisions at \sqrts\ = 200 GeV in both the Helicity and Collins-Soper reference frames. Results utilizing the dimuon and dielectron decay channels are presented and agree with each other within uncertainties although slightly different kinematic ranges are covered. The inclusive \jpsi's do not exhibit significant transverse or longitudinal polarization with little dependence on \pT. Among several model calculations compared to data, the CGC+NRQCD agrees the best overall. These results provide additional tests and valuable guidance for theoretical efforts towards a complete understanding of the \jpsi\ production mechanism in vacuum.

\section*{Acknowledgements}
We thank the RHIC Operations Group and RCF at BNL, the NERSC Center at LBNL, and the Open Science Grid consortium for providing resources and support.  This work was supported in part by the Office of Nuclear Physics within the U.S. DOE Office of Science, the U.S. National Science Foundation, the Ministry of Education and Science of the Russian Federation, National Natural Science Foundation of China, Chinese Academy of Science, the Ministry of Science and Technology of China and the Chinese Ministry of Education, the Higher Education Sprout Project by Ministry of Education at NCKU, the National Research Foundation of Korea, Czech Science Foundation and Ministry of Education, Youth and Sports of the Czech Republic, Hungarian National Research, Development and Innovation Office, New National Excellency Programme of the Hungarian Ministry of Human Capacities, Department of Atomic Energy and Department of Science and Technology of the Government of India, the National Science Centre of Poland, the Ministry  of Science, Education and Sports of the Republic of Croatia, RosAtom of Russia and German Bundesministerium fur Bildung, Wissenschaft, Forschung and Technologie (BMBF), Helmholtz Association, Ministry of Education, Culture, Sports, Science, and Technology (MEXT) and Japan Society for the Promotion of Science (JSPS).


\section*{Appendix: Data tables}
\label{sect:app}
The values of inclusive $J\psi$ polarization parameters in different \pT\ bins are shown in Tables \ref{Final_stas_HX_EE}, \ref{Final_stas_CS_EE}, \ref{Final_stas_HX_MuMu}, and \ref{Final_stas_CS_MuMu}.

\begin{table*}[htbp]
\renewcommand\arraystretch{1.3} 
\centering
\caption{The inclusive \jpsi\ polarization parameters in the HX frame in different \pT\ bins measured through the dielectron channel within $|y| < 1$. The first uncertainty is statistical and the second is systematic.}
\label{Final_stas_HX_EE}
\newcolumntype{V}{!{\vrule width 2.0pt}}
\begin{tabular}{|c |c | r @{$\pm$} @{} r |  r @{$\pm$} @{} r  |  r @{$\pm$} @{} r  | r @{$\pm$} @{} r  | r @{$\pm$} @{} r  | r @{$\pm$} @{} r |  r @{$\pm$} @{} r  | r @{$\pm$} @{} r | }
\hline
\pT\ (\gev) & $\langle \pT \rangle$ (\gev)& \multicolumn{2}{c|}{$\uplambda_{\uptheta}^{HX}$}  & \multicolumn{2}{c|}{$\uplambda_{\upphi}^{HX}$} & \multicolumn{2}{c|}{$\uplambda_{\uptheta\upphi}^{HX}$} & \multicolumn{2}{c|}{$\uplambda_{\rm{inv}}^{HX}$} \\ \hline
0-2   & 1.08 &  0.22&0.46$\pm$0.19 &  0.11&0.12$\pm$0.10 &  0.20&0.23$\pm$0.12 &  0.63&0.63$\pm$0.36 \\ \hline
2-4   & 2.69 & -0.06&0.19$\pm$0.14 &  -0.04&0.17$\pm$0.09 &  0.02&0.07$\pm$0.05 &  -0.16&0.39$\pm$0.30  \\ \hline
4-6   & 4.66 & -0.03&0.43$\pm$0.29 &  -0.28&0.41$\pm$0.22 &  0.06&0.09$\pm$0.05 &  -0.68&0.66$\pm$0.38  \\ \hline
6-8 & 6.68 &   0.01&0.26$\pm$0.30 & -0.03&0.12$\pm$0.07 & 0.12&0.12$\pm$0.07 &  -0.07&0.38$\pm$0.26 \\ \hline
8-14 & 9.15&  0.95&0.91$\pm$0.37 & -0.21&0.28$\pm$0.08 & 0.27&0.35$\pm$0.22 &  0.25&1.05$\pm$0.38  \\ \hline
\end{tabular}
\end{table*}

\begin{table*}[htbp]
\renewcommand\arraystretch{1.3} 
\centering
\caption{The inclusive \jpsi\ polarization parameters in the CS frame in different \pT\ bins measured through the dielectron channel within $|y| < 1$. The first uncertainty is statistical and the second systematic.}
\label{Final_stas_CS_EE}
\newcolumntype{V}{!{\vrule width 2.0pt}}
\begin{tabular}{|c |c | r @{$\pm$} @{} r |  r @{$\pm$} @{} r  |  r @{$\pm$} @{} r  | r @{$\pm$} @{} r  | r @{$\pm$} @{} r  | r @{$\pm$} @{} r |  r @{$\pm$} @{} r  | r @{$\pm$} @{} r | }
\hline
\pT\ (\gev) & $\langle \pT \rangle$ (\gev) & \multicolumn{2}{c|}{$\uplambda_{\uptheta}^{CS}$}  & \multicolumn{2}{c|}{$\uplambda_{\upphi}^{CS}$} & \multicolumn{2}{c|}{$\uplambda_{\uptheta\upphi}^{CS}$} & \multicolumn{2}{c|}{$\uplambda_{\rm{inv}}^{CS}$} \\ \hline
0-2   &  1.08 & 0.78&1.01$\pm$0.45 &   0.16&0.16$\pm$0.08 &  -0.24&0.29$\pm$0.12 &  1.49&1.57$\pm$0.92 \\ \hline
2-4   & 2.69 & -0.46&0.35$\pm$0.21 &  -0.09&0.08$\pm$0.08 &  -0.04&0.11$\pm$0.05 &  -0.67&0.38$\pm$0.26  \\ \hline
4-6   & 4.66 & -0.25&0.36$\pm$0.22 &  -0.04&0.10$\pm$0.12 &  0.04&0.12$\pm$0.11 &  -0.35&0.46$\pm$0.32  \\ \hline
6-8 & 6.68 &  -0.25&0.22$\pm$0.15 &    0.03&0.12$\pm$0.18 & -0.12&0.11$\pm$0.06 &  -0.17&0.41$\pm$0.36  \\ \hline
8-14 &  9.15 &  -0.69&0.22$\pm$0.07 & 0.18&0.20$\pm$0.12 & -0.07&0.19$\pm$0.15 &  -0.18&0.74$\pm$0.35  \\ \hline
\end{tabular}
\end{table*}

\begin{table*}[htbp]
\renewcommand\arraystretch{1.3} 
\centering
\caption{The inclusive \jpsi\ polarization parameters in the HX frame in different \pT\ bins measured through the dimuon channel within $|y| < 0.5$. The first uncertainty is statistical and the second systematic.}
\label{Final_stas_HX_MuMu}
\newcolumntype{V}{!{\vrule width 2.0pt}}
\begin{tabular}{|c |c | r @{$\pm$} @{} r |  r @{$\pm$} @{} r  |  r @{$\pm$} @{} r  | r @{$\pm$} @{} r  | r @{$\pm$} @{} r  | r @{$\pm$} @{} r | }
\hline
\pT\ (\gev) & $\langle \pT \rangle$ (\gev) & \multicolumn{2}{c|}{$\uplambda_{\uptheta}^{HX}$}  & \multicolumn{2}{c|}{$\uplambda_{\upphi}^{HX}$} & \multicolumn{2}{c|}{$\uplambda_{\rm{inv}}^{HX}$} \\ \hline
0-1   & 0.62 & -0.01&0.15$\pm$0.11 &  -0.04&0.06$\pm$0.03 &  -0.12&0.22$\pm$0.13 \\ \hline
1-2   & 1.46 & -0.34&0.32$\pm$0.31 &  0.15&0.25$\pm$0.09  &  0.14&0.99$\pm$0.54  \\ \hline
2-4   & 2.69 & -0.18&0.22$\pm$0.11 &  0.29&0.09$\pm$0.10  &  0.98&0.60$\pm$0.40  \\ \hline
4-10 & 4.92 &  0.12&0.42$\pm$0.17 & -0.07&0.19$\pm$0.07  &  0.35&0.82$\pm$0.30 \\ \hline
\end{tabular}
\end{table*}

\begin{table*}[htbp]
\renewcommand\arraystretch{1.3} 
\centering
\caption{The inclusive \jpsi\ polarization parameters in the CS frame in different \pT\ bins measured through the dimuon channel within $|y| < 0.5$. The first uncertainty is statistical and the second systematic.}
\label{Final_stas_CS_MuMu}
\newcolumntype{V}{!{\vrule width 2.0pt}}
\begin{tabular}{|c |c | r @{$\pm$} @{} r |  r @{$\pm$} @{} r  |  r @{$\pm$} @{} r  | r @{$\pm$} @{} r  | r @{$\pm$} @{} r  | r @{$\pm$} @{} r | }
\hline
\pT\ (\gev) & $\langle \pT \rangle$ (\gev) & \multicolumn{2}{c|}{$\uplambda_{\uptheta}^{CS}$}  & \multicolumn{2}{c|}{$\uplambda_{\upphi}^{CS}$} & \multicolumn{2}{c|}{$\uplambda_{\rm{inv}}^{CS}$} \\ \hline
0-1   & 0.62 & -0.23&0.89$\pm$0.32 &   -0.03&0.05$\pm$0.04  &   -0.30&0.88$\pm$0.34 \\ \hline
1-2   & 1.46 &  0.54&0.71$\pm$0.45 &   0.05&0.18$\pm$0.11   &   0.72&1.02$\pm$0.51  \\ \hline
2-4   & 2.69 &  0.66&0.45$\pm$0.15 &  -0.11&0.19$\pm$0.10   &   0.30&0.69$\pm$0.25  \\ \hline
4-10 & 4.92 & -0.02&0.36$\pm$0.15 &  -0.33&0.24$\pm$0.22  &  -0.76&0.48$\pm$0.35  \\ \hline
\end{tabular}
\end{table*}

\end{document}